\documentclass[11pt,a4paper]{article}
\usepackage{jheppub}
\usepackage{amsthm,bm,mathrsfs}

\newcommand{\be}{\begin{equation}}
\newcommand{\ee}{\end{equation}}
\newcommand{\ba}{\begin{eqnarray}}
\newcommand{\ea}{\end{eqnarray}}

\newcommand{\m}[1]{$\mathop{#1}$}

\newcommand{\la}{\langle}
\newcommand{\ra}{\rangle}

\newcommand{\ppd}[2]{\big<\!\!\big<\,{#1}\,\big|\,{#2}\,\big>\!\!\big>}

\newcommand{\st}{\scriptstyle}
\newcommand{\sst}{\scriptscriptstyle}

\newcommand{\nn}{\nonumber\\}
\newcommand{\e}{&=&}

\author[a]{Euihun Joung}
\author[b]{Jihad Mourad}

\affiliation[a]{Scuola Normale Superiore and INFN\\
Piazza dei Cavalieri 7, 56126 Pisa, Italy}
\affiliation[b]{APC\footnote{ Unit\'e Mixte de Recherche $7164$ du
CNRS}, Universit\'e Paris VII,\\ B\^atiment Condorcet, 75205 Paris
Cedex 13, France}

\emailAdd{euihun.joung@sns.it}
\emailAdd{mourad@apc.univ-paris7.fr}

\title{\center{Boundary action  of \\
 free AdS higher-spin gauge fields\\
 and the holographic correspondence}}

\abstract{We determine the boundary terms of the free higher-spin
action which reproduce the AdS Fronsdal equations in an AdS
manifold with a finite distance boundary. The boundary terms are
further constrained by the gauge invariance of the total action. We show
that, for spins larger than two, no local boundary term can restore the
full gauge symmetry, and the broken symmetry corresponds to higher-spin Weyl
transformations on the boundary CFT.
 The boundary action is used for the evaluation of the
on-shell higher-spin AdS action in terms of the boundary data
given by a conformal higher-spin field.}

\begin{document}

\maketitle

\section{Introduction}

One of the most remarkable holographic correspondences is the one
relating an interacting higher-spin (HS) theory\footnote{ For
recent reviews on higher-spin theory, see e.g. the proceeding
\cite{SolvayHS} (which includes the contributions \cite{
Bianchi:2005yh, Francia:2005bv, Bouatta:2004kk, Bekaert:2005vh,
Sagnotti:2005ns}) and \cite{Sorokin:2004ie,
Francia:2006hp,Bekaert:2010hw,Sagnotti:2011qp}.
For recent works on massless higher-spin fields from string theory, see e.g.
\cite{Bonelli:2003zu,Bonelli:2004ve,Polyakov:2009pk,Polyakov:2010qs,Polyakov:2010sk,Polyakov:2011sm,Sagnotti:2010at}.
} in Anti de Sitter
spacetime (AdS) to a boundary free scalar theory
\cite{Sundborg:2000wp, Witten:0111xxx,Mikhailov:2002bp,
Tseytlin:2002gz,Sezgin:2002rt,Segal:2002gd,Klebanov:2002ja,Maldacena:2011jn}.
This conjectured
correspondence allows, in principle, to connect the observables of
the HS theory to observables in the free CFT coupled to external
HS sources \cite{Witten:1998qj}. Since the CFT is free, one
expects to deduce many features of the HS theory and to gain a
better understanding of the Vasiliev system (see e.g.
\cite{Vasiliev:2004cp,Vasiliev:2004qz} for some reviews), the
candidate of the AdS theory. Despite many checks
\cite{Mikhailov:2002bp,Girardello:2002pp,Sezgin:2003pt,
Petkou:2003zz, Petkou:2004nu,
Manvelyan:2004ii,Manvelyan:2005fp,Manvelyan:2005ew,Manvelyan:2006zy,
Manvelyan:2008ks, Metsaev:2008fs, Metsaev:2009ym,
Giombi:2009wh,Giombi:2010vg} (see also
\cite{Campoleoni:2010zq,Henneaux:2010xg} for the
AdS$_{3}$/CFT$_{2}$ case), the state of affairs is still, in our
opinion, unsatisfactory because a systematic comparison is
lacking.

In this article we propose to initiate a systematic study starting from the
reasonable assumption that the HS theory can be described
as a series in the coupling constant  with the first term being
the free Fronsdal action \cite{Fronsdal:1978vb}.
 The subject of this paper is the
determination of the free on-shell Fronsdal action as a functional
of the boundary data. This is the object to be compared with the
quadratic part of the CFT effective action.

The  AdS/CFT correspondence applied to gauge fields
in the semi-classical regime states that the finite part of
on-shell AdS action  coincides with the generating functional for
connected correlation functions of conserved currents on the
boundary. These two quantities are to be matched after the
identification of the asymptotic behavior of the bulk gauge fields
with the source fields of the conserved currents. In the case of
the HS holography, the boundary behaviors of an infinite number of
HS gauge fields in AdS turn into the sources of an infinite tower
of conserved currents of higher ranks. These currents, for free
scalar fields,  are bilinear  so that the generating functional
can be viewed as the finite part of the quantum effective action
of the scalar field in an external HS background
\cite{Bekaert:2010ky}.

The ultra-violet divergences of the effective action can be
handled with a cut-off and holographically they become infra-red
divergences  regularized by displacing the boundary of AdS to a
finite distance. 
As a consequence,  the AdS Fronsdal action needs
to be supplemented by boundary terms. In the presence of a
boundary,
 the variation of the AdS Fronsdal action leads to
boundary terms whose cancellation determines the HS generalization
of the spin two York-Gibbons-Hawking terms
\cite{York:1972sj,Gibbons:1976ue}. This is the subject of Section
\ref{sec: bd actions}. The boundary terms have some arbitrariness
because the addition of boundary terms independent of the radial
derivative does not change the equations of motions. In Section
\ref{sec: gen bd}, we constrain these terms in order to keep as
much as possible of the gauge symmetries of the action. We show
that it is not possible  to restore all the bulk gauge symmetries
with local counterterms,
and identify the broken part of the symmetries.
The latter generalizes to HS the radial diffeomorphism of Gravity.
This broken gauge symmetry lead in general to
the corresponding anomaly of the AdS on-shell action,
and should match with the HS Weyl anomaly of CFT
 \cite{Manvelyan:2006zy,Bekaert:2010ky}.\footnote{
 See \cite{Henningson:1998gx} for the Gravity case.}
We then turn to the evaluation of the on-shell action
as a functional of  the boundary 
behavior of spin $s$ Fronsdal field.
The latter actually contains more independent components than
needed to couple to a rank $s$ conserved current. But we show that
these components are related on-shell to each others and reduce to
a $d$-dimensional spin $s$ field.
This field, which we call a HS conformal field, is not subject to
trace constraints, but as the classical CFT action suggests, has a
HS generalization of gauge and Weyl symmetries.\footnote{ The
conformal HS fields were considered in \cite{Fradkin:1985am,
Fradkin:1989md,Fradkin:1990re, Segal:2002gd}. } The HS Weyl
symmetry of the boundary theory is in general broken by quantum
corrections. The holographic dual of this is developed in Section
\ref{sec: OA} where we deduce the expression of the on-shell
action in terms of the boundary data provided by the
$d$-dimensional conformal
spin $s$ field. For spin two, we show how  
one can obtain
the same answer as the more conventional boundary field given
by the bulk field in
radial gauge evaluated on the boundary.
In the general case, we give the expression of the finite part of
the on-shell action and show that it is identical, up to an
overall constant, to the finite part of the effective action as
determined in \cite{Bekaert:2010ky}.
This is due to the cancellation of the bulk anomaly
by the local counterterm identified in Section  \ref{sec: gen bd}.
Finally, we collect our conclusions in Section
\ref{sec: concl} 
where, we discuss our results in the light of the
AdS/CFT correspondence and in particular, we compare the on shell
action with the quadratic part of the CFT effective action
confirming the HS holography conjecture at this level.
 The technical details used in the text can be found in the appendices.

\section{Boundary action}
\label{sec: bd actions}

\subsection{Bulk action and notations}

We first recall the AdS Fronsdal HS equations and action
in order to explicit our notations.
 We adopt the Poincar\'e coordinate system where the
$(d+1)$-dimensional AdS is parametrized by $x^{\sst M}\in
\mathbb{R}^{d}\times \mathbb{R}_{>0}$\,. The AdS background metric
$\bar g_{\sst MN}$ is given by \ba
    ds^{2} = \bar g_{\sst MN}\,d x^{\sst M}\, dx^{\sst N}
    \e \frac{1}{{(x^{d})}^{2}}\,\eta_{\sst MN}\,d x^{\sst M}\, dx^{\sst N}
    \hspace{48pt} [{\st M}, {\st N}=0,1,\cdots,d\,] \nn
    \e \frac{1}{z^{2}} \left(\eta_{\mu\nu}\,d x^{\mu}\,d x^{\nu}+dz^{2}\right)
    \qquad [\mu,\nu=0,1,\cdots,d-1]\,,
\ea where $z:= x^{d}$\,. The constant-$z$ hyperplanes are
$d$-dimensional Minkowski spaces, and the $z=0$ surface is the
boundary of AdS. The regularization of the on-shell action is
performed by transporting the   boundary to $z_B\neq 0$\,.

It is convenient, in order to simplify the notations, to handle
the tensor fields by contracting them with auxiliary variables.
More precisely in the spin $s$ case, we contract the symmetric
tensor $\varphi_{\sst M_1\dots M_s}$ with auxiliary variables
$U^{\sst A}\in \mathbb{R}^{d+1}$\,, by making use of the vielbein
of the AdS metric: $\bar e_{\sst A}^{\ \sst
M}(x,z)=z\,\delta^{\sst M}_{\sst A}$\,, as \be
    \varphi(x,z;U) := \tfrac{1}{s!}\,U^{\sst A_1}\,\dots U^{\sst A_s}\
    \bar e_{\sst A_1}^{\sst \ \, M_1}(x,z)\ \dots \bar e_{\sst A_s}^{\sst\ \, M_s}(x,z)\
    \varphi_{\sst M_1\dots M_s}(x,z)\,.
\ee With this notation, the doubly traceless constraint reads
$(\partial_{U}^{2})^{2}\,\varphi=0$\, while the
equations of motion (EOM) of the massless spin $s$ field  $F_{\sst
M_1\dots M_s}\approx 0$ can be expressed  as \be
    \tfrac{1}{s!}\,U^{\sst A_1}\,\dots U^{\sst A_s}\,
    \bar e_{\sst A_1}^{\sst \ \, M_1}\,\dots  \bar e_{\sst A_s}^{\sst\ \, M_s}\,
    F_{\sst M_1\dots M_s}=
    \mathcal{F}\,\varphi(U)\,,
\ee where the Fronsdal  operator $\mathcal F$ acts on the
auxiliary variables as well as the spacetime variables
 with the AdS covariant
expression: \be
    \mathcal{F}=\Box_{AdS}-(U\cdot \mathcal{D})
    (\partial_{U}\cdot \mathcal{D})
    +\tfrac12\,(U\cdot \mathcal{D})^{2}\,\partial_{U}^{2}-U^{2}\,\partial_{U}^{2}
    -m_{s}^{2}\,,
    \label{Fronsdal op}
\ee where $\mathcal{D}$ and $\Box_{AdS}$ are respectively the AdS
covariant derivative and the D'Alembertian operator in auxiliary
variables:
 \ba
    \mathcal{D}_{A} \e \bar e_{\sst A}^{\sst \ \, M}\, \partial_{x^{\sst M}}
    + \tfrac12\,\bar \omega^{\ \ \, \sst C}_{\sst AB}\,
    U^{\sst B}\, {\partial_{U^{\sst C}}}=
    z\,\partial_{x^{\sst A}}
    +U^{d}\,\partial_{U^{\sst A}}-U^{\sst A}\,\partial_{U^{d}}\,, \nn
    \square_{AdS} \e \partial_{U'}\cdot(\mathcal{D}
    +U'^{d}\,\partial_{U'}-U'\,\partial_{U'^{d}})\,
    (U'\cdot \mathcal{D})\,,
\ea
 and $m_{s}^{2}$ is given by
\be
    m^{2}_{s} = s^{2}+(d-5)\,s-2(d-2)\,.
    \label{Fr mass}
\ee Using these notations for the  fields and EOM, we now express
the action  of the free massless spin $s$ field in
the region $z\ge z_{\sst B}$ of $AdS_{d+1}$ denoted in the following by
$\mathcal{M}$:
\be
    \mathcal{I}_{\sst \mathcal M}
    =-\frac12 \int^{\infty}_{z_{B}} \frac{dz}{z^{d+1}}\,
    \ppd{\varphi}{
    \left(1-\tfrac14\,U^{2}\,\partial_{U}^{2} \right) \mathcal{F}\,\varphi}\,,
    \label{Frons act}
\ee where the bracket $\ppd{\cdot}{\cdot}$ is defined as \be
    \ppd{f}{g}:=\int d^{d}x\,\sum_{n=0}^{\infty}
    \frac1{n!}\,f_{\mu_{1}\cdots\mu_{n}}(x)\,g^{\mu_{1}\cdots\mu_{n}}(x)\,.
\ee This Fronsdal action is our starting point to construct the
boundary terms which allow to derive the  EOM from the variational
principle and  which insure that the total action is invariant
under the largest subset of gauge transformations. In fact, the
Fronsdal equation $\mathcal F\,\varphi\approx 0$ is
invariant under the gauge transformation: \be
    \delta_{\varepsilon}\,\varphi(x,z;U)
    =U\cdot \mathcal D\ \varepsilon(x,z;U)\,,
    \label{tr s}
\ee where the gauge parameter $\varepsilon$ is
subject to the traceless constraint: \be
    \partial_{U}^{2}\,\varepsilon(x,z;U)=0\,.
\ee The gauge invariance of the Fronsdal action in AdS is due to
the Bianchi identity: \be
    (\partial_{U}\cdot \mathcal{D}-\tfrac12\,U\cdot\mathcal D\,\partial_{U}^{2})\,
    \mathcal{F}=0\,.
    \label{Bianchi}
\ee

\subsection{Radial decomposition}

 In order to unravel the field content from the boundary point of view,
  we decompose the $O(d,1)$-tensors in terms of the boundary
$O(d)$-tensors. This is conveniently done by
denoting $U^{\sst A}$ as $(u^{\mu},v)$ and
by expanding the fields
$\varphi(u,v)$ and the gauge parameters
$\varepsilon(u,v)$  in powers of $v$ as
\ba
\label{phi decomp}
   && \varphi(x,z;u,v) := \sum_{r=0}^{s}\,\frac{v^{r}}{r!}\,
    \phi^{\sst(s-r)}(x,z;u)\,, \\
\label{eps decomp}
    && \varepsilon(x,z;u,v) :=
    \sum_{r=0}^{s-1}\,\frac{v^{r}}{r!}\,
    \epsilon^{\sst(s-1-r)}(x,z;u)\,.
\ea
The doubly tracelessness and the tracelessness constraints
allow to express $\phi^{\sst (r)}$ for $r=0,\cdots, s-4$ and
$\epsilon^{\sst (r)}$ for $r=0,\cdots, s-3$ as \ba
    && \phi^{\sst (s-2n-m)}  = (-1)^{n+1}
    \left[ (n-1)\,(\partial_u^{2})^{n} \phi^{\sst (s-m)}
    + n\,(\partial_u^{2})^{n-1} \phi^{\sst (s-2-m)} \right],
    \label{decomp. fields} \\
     && \epsilon^{\sst (s-1-2n-m)}  =
    (-1)^{n} (\partial_u^{2})^{n} \epsilon^{\sst (s-1-m)},
\ea where $n=1,\cdots $ and $m=0$ or $1$. Thus the independent
gauge fields and parameters are the unconstrained
\be
    \phi^{\sst (s)}\,, \qquad
    \phi^{\sst (s-1)}\,, \qquad
    \phi^{\sst (s-2)}\,, \qquad
    \phi^{\sst (s-3)}\,,
\ee and \be
    \epsilon^{\sst (s-1)}\,, \qquad  \epsilon^{\sst (s-2)}\,.
\ee In terms of these $O(d)$-tensors, the gauge transformation
\eqref{tr s} reads \ba
    && \delta\,\phi^{\sst(s)}
    =z\,(u\cdot\partial )\,\epsilon^{\sst (s-1)}
    -u^{2}\,\epsilon^{\sst (s-2)}\,,\nn
    && \delta\,\phi^{\sst(s-1)}
    =(z\,\partial_{z}+s-1+u^{2}\,\partial_{u}^{2})\,
    \epsilon^{\sst (s-1)}
    +z\,(u\cdot\partial )\,\epsilon^{\sst (s-2)}\,,\nn
    && \delta\,\phi^{\sst(s-2)}
    =-z\,(u\cdot\partial )\,\partial_{u}^{2}\,\epsilon^{\sst (s-1)}
    +\left[2\left(z\,\partial_{z}+s-2\right)+u^{2}\,\partial_{u}^{2}\right]
    \epsilon^{\sst(s-2)}\,,\nn
    && \delta\,\phi^{\sst(s-3)}
    =-\left[3\left(z\,\partial_{z}+s-3\right)+u^{2}\,\partial^{2}_{u}\right]
    \partial_{u}^{2}\,\epsilon^{\sst (s-1)}
    -z\,(u\cdot\partial )\,\partial_{u}^{2}\,\epsilon^{\sst (s-2)}\,.
    \label{decomp. bd necessaries}
\ea Notice that the transformation of $\phi^{\sst (s)}$ is a
combination of a $d$-dimensional HS gauge transformation
with an unconstrained parameter and a
HS Weyl transformation characterizing a conformal $d$-dimensional gauge
field. From the boundary point of view at constant $z$ there are
two spin $s-1$ unconstrained gauge parameters $\epsilon^{\sst
(s-1)}$ and $\partial_z\,\epsilon^{\sst (s-1)}$ and two
spin $s-2$ unconstrained gauge parameters $\epsilon^{\sst
(s-2)}$ and $\partial_z\,\epsilon^{\sst (s-2)}$. Notice
also that $\partial_z\,\epsilon^{\sst (s-1)}$ and
$\partial_z\,\epsilon^{\sst (s-2)}$ can be seen as
Stueckelberg shift fields for $\phi^{\sst(s-1)}$ and
$\phi^{\sst(s-2)}$. The latter pair of fields can thus be set
to zero by fixing $\partial_z\,\epsilon^{\sst (s-1)}$ and
$\partial_z\,\epsilon^{\sst (s-2)}$.\footnote{
In eq.~\eqref{decomp. bd necessaries} and in the following, we will use the
notation $\partial=\partial/(\partial x)$ and the derivative with respect to
$z$ will be explicitly spelled $\partial_z$.}

\subsection{Construction of boundary action}

In the presence of a finite distance boundary, the bulk action
$\mathcal{I}_{\sst\mathcal M}$ \eqref{Frons act} is not sufficient
to get the EOM from the variational principle.
This can be easily seen by calculating  the on-shell  variation of the
bulk action $\delta\,\mathcal{I}_{\sst\mathcal M}$ under
$\delta\,\varphi$\,: \be
    \delta\,\mathcal{I}_{\sst\mathcal M}
    \approx-\frac12\int^{\infty}_{z_{B}} \frac{dz}{z^{d+1}}\,
    \ppd{ \varphi}{
    \left(1-\tfrac14\,U^{2}\,\partial_{U}^{2} \right) \mathcal{F}\,
    \delta\,\varphi},
    \label{spin s var}
\ee where we have used the EOM. If $\mathcal M$ were a manifold
without boundary, then integrating by part \eqref{spin s var},
$\delta\,\mathcal{I}_{\sst \mathcal M}$ could be expressed with
the EOM and so would vanish on-shell. But for the present case,
$\mathcal M$ has a boundary at $z=z_{B}$ and the integration by
part results in an additional  boundary term. Remember also that
when varying the action we should use a proper boundary condition,
either the Dirichlet condition $\delta\,\varphi(z_{B})=0$ or the Neumann condition
$\delta\,\partial_{z}\varphi(z_{B})=0$ (or a mixed
one). Following the standard version of AdS/CFT, we use the
Dirichlet boundary condition. By integrating \eqref{spin s var} by
parts and by using $\delta\,\varphi(z_{B})=0$ and the
EOM, we obtain $\delta\,\mathcal{I}_{\sst \mathcal M}$ as a
boundary term: \be
    \delta\,\mathcal{I}_{\sst \mathcal M}\approx
    \tfrac{z_{B}^{-d-1}}2\,
    \ppd{\varphi}{\left(1-\tfrac14\,U^{2}\,\partial_{U}^{2} \right)
     \mathcal{F}_{zz}\,\partial_{z}\,\delta\,\varphi}_{z_{B}}\,,
     \label{var spin 2}
\ee where $\mathcal F_{zz}$ is the coefficient of
$\partial_{z}^{2}$ in the Fronsdal operator \eqref{Fronsdal op}.
Explicitly it is given
 by \be
    \mathcal{F}_{zz}=z^{2}\left(1-v\,\partial_{v}+\tfrac12\,v^{2}\,\partial_{U}^{2}\right).
    \label{Fzz}
\ee We see that the on-shell variation $\delta\,\mathcal I_{\sst
\mathcal M}$ does not vanish, and this implies that the
variational principle does not  lead to  the EOM and needs the
addition of a boundary term $\mathcal I^{\sst\,0}_{\sst
\partial \mathcal M}$
so that the total action satisfies \be
    \delta\left(\,\mathcal I_{\sst \mathcal M}
    +\mathcal I^{\sst\,0}_{\sst \partial \mathcal M}\,
    \right)\approx 0\,.
\ee

In order that the bulk action provides the Fronsdal equation from
the variational principle with $\delta\,\varphi(z_{B})=0$\,,
the boundary term
$\mathcal{I}_{\sst\partial \mathcal M}^{\sst\,0}$ is determined by
\be
    \mathcal{I}_{\sst\partial \mathcal M}^{\sst\,0}=
    -\tfrac{z_{B}^{-d-1}}2\,
    \ppd{\varphi}{
    \left(1-\tfrac14\,U^{2}\,\partial_{U}^{2} \right)
     \mathcal{F}_{zz}\,\partial_{z}\,\varphi}_{z_{B}}\,.
     \label{bd ss}
\ee
By using the
radial decomposition of the field $\varphi$, we obtain
the explicit expression of \eqref{bd ss} in terms of $O(d)$-tensor
fields as
\be
     \mathcal{I}_{\sst \partial \mathcal M}^{\sst\,0}=
     \tfrac{z_{B}^{-d}}2 \left[
    \ppd{ \phi^{\sst (s)}}{\mathsf{P}_{\rm e}'\,
    z\,\partial_{z}\,\phi^{\sst (s)}}_{z_{B}}+
    \ppd{ \psi^{\sst (s-3)}}{
    \mathsf{P}_{\rm o}'\,z\,\partial_{z}\,\psi^{\sst (s-3)}}_{z_{B}} \right],
    \label{bd. term exp}
\ee with \be
    \psi^{\sst (s-3)} := \tfrac 12
     \left(\phi^{\sst (s-3)}+3\,\partial_{u}^{2}\,\phi^{\sst (s-1)}\right),
\ee and \be
    \mathsf{P}_{\rm e}':=\sum_{n=0}^{\infty}
    \tfrac{2n-1}{(2\,n)!}\,(u^{2})^{n}(\partial_{u}^{2})^{n}\,,
     \qquad
     \mathsf{P}_{\rm o}':= \sum_{n=0}^{\infty}
    \tfrac{2n+2}{(2\,n+3)!}\,(u^{2})^{n}(\partial_{u}^{2})^{n}\,.
\ee Notice that the boundary term $ \mathcal{I}_{\sst \partial
\mathcal M}^{\sst\,0}$ does not depend on $\phi^{\sst
(s-2)}$, and it depends on $\phi^{\sst (s-1)}$ and
$\phi^{\sst (s-3)}$ only through $\psi^{\sst (s-3)}$.

The action
$\mathcal{I}_{\sst\mathcal M}+\mathcal{I}^{\sst\,0}_{\sst\partial
\mathcal M}$ provides the Fronsdal equation in $\mathcal M$, but,
since we are using a Dirichlet boundary condition, any boundary
term which does not involve $\partial_z\,\phi^{\sst (r)}$ can be
freely added without affecting the variational principle. This
gives rise to an ambiguity on the on-shell action and consequently in the AdS/CFT
correspondence. In order to fix the ambiguity, we
rely on the variation under the gauge transformation of the total action:
\be
    \delta_{\varepsilon}\left(\,
    \mathcal{I}_{\sst\mathcal M} +
    \mathcal{I}_{\sst\partial \mathcal M}\,\right)=0\,,
\ee where $\mathcal{I}_{\sst\partial \mathcal M}$ is a completion
of $\mathcal{I}_{\sst\partial \mathcal M}^{\sst\,0}$ with
$\partial_z\phi^{\sst (r)}$-free boundary terms. For the
determination of $\mathcal{I}_{\sst\partial \mathcal M}$, we
proceed in two steps:
\begin{enumerate}
\item we determine the gauge variation of $\mathcal{I}_{\sst\mathcal
M} + \mathcal{I}^{\sst\,0}_{\sst\partial \mathcal M}$ under
the transformations $\alpha$ generated by $\epsilon^{\sst (s-1)}, \partial_{z}\epsilon^{\sst (s-1)}$
and $\partial_{z}\epsilon^{\sst (s-2)}$\,, and then
add a $\partial_z\phi^{\sst (r)}$-independent boundary term such that
\be
    \delta_{\alpha}\left(\,
    \mathcal{I}_{\sst\mathcal M} + \mathcal{I}_{\sst\partial \mathcal M}\,\right)=0\,;
    \label{2 bd}
\ee
The result of this step is summarised in
eqs.~(\ref{boundary}) and (\ref{defs}).
\item The most general boundary term is then given by the
addition to the above term of a $\partial_z\phi^{\sst (r)}$-independent
local term invariant under the above
transformations $\alpha$. We determine these counterterms and
examine whether they can compensate the
gauge variation of $\mathcal{I}_{\sst
\mathcal M} + \mathcal{I}_{\sst \partial \mathcal M}$ under
$\epsilon^{\sst (s-2)}$\,.
The outcome of this step are the counterterms (\ref{c}) and
(\ref{c.t. 2}) with the definitions (\ref{def A C}) and \eqref{GK}.
We will also prove in Section  \ref{subsec: Anomaly} that,
for spin larger than two, no counterterm can fully compensate the $\epsilon^{\sst (s-2)}$
variation and hence the action is characterized by an
anomaly under the HS Weyl transformations.
\end{enumerate}

\smallskip

Let us first specialize to the spin 2 case for which the formulas
simplify considerably.

\subsection{Spin 2 case}
The boundary term \eqref{bd. term exp} reduces to
\be
    \mathcal{I}_{\sst \partial \mathcal M}^{\sst\,0}=
   -\tfrac{z_{B}^{-d}}2\,
    \ppd{ \phi^{\sst (2)}}{
    \left(1-\tfrac 12\,u^{2}\,\partial_{u}^{2} \right)
    z\,\partial_{z}\,\phi^{\sst (2)}}_{z_{B}}.
    \label{var 2}
\ee
Any boundary term independent of $\partial_{z}\,\phi^{\sst
(r)}$ vanishes under the variation $\delta\,\varphi$
with the Dirichlet condition, and can be freely added without
effecting the EOM.

In order to fix this ambiguity of boundary terms, as stated above, we will rely on
the gauge invariance. The linearized diffeorphism
is expressed in auxiliary variables in eq.~\eqref{tr s}, and
in terms of $\phi^{\sst (2)}, \phi^{\sst (1)}, \phi^{\sst (0)}$ \eqref{phi decomp} and
$\epsilon^{\sst (1)}, \epsilon^{\sst (0)}$ \eqref{eps decomp},  the gauge
transformation \eqref{decomp. bd necessaries} reads
\ba
    && \delta\,\phi^{\sst(2)}
    =z\,(u\cdot\partial )\,\epsilon^{\sst (1)}
    -u^{2}\,\epsilon^{\sst (0)}\,,\nn
    && \delta\,\phi^{\sst(1)}
    =(z\,\partial_{z}+1)\,
    \epsilon^{\sst (1)}
    +z\,(u\cdot\partial )\,\epsilon^{\sst (0)}\,,\nn
    && \delta\,\phi^{\sst(0)}
    =2\,z\,\partial_{z}\,\epsilon^{\sst(0)}\,.
    \label{spin 2 gt}
\ea
Notice that the transformation of $\phi^{\sst (2)}$ is a
combination of a $d$-dimensional gauge transformation and a Weyl
transformation characterizing a conformal $d$-dimensional spin 2 gauge
field. From the boundary point of view at constant $z$ there are
two spin 1 gauge parameters $\epsilon^{\sst
(1)}(z_{B})$ and $\partial_z\,\epsilon^{\sst (1)}(z_{B})$ and two
spin 0 gauge parameters $\epsilon^{\sst
(0)}(z_{B})$ and $\partial_z\,\epsilon^{\sst (0)}(z_{B})$.
Notice also that $\partial_z\,\epsilon^{\sst (1)}(z_{B})$ and
$\partial_z\,\epsilon^{\sst (0)}(z_{B})$ can be seen as
Stueckelberg shift fields for $\phi^{\sst(1)}(z_{B})$ and
$\phi^{\sst(0)}(z_{B})$\,. The latter pair of fields can thus be set
to zero by fixing $\partial_z\,\epsilon^{\sst (1)}(z_{B})$ and
$\partial_z\,\epsilon^{\sst (0)}(z_{B})$\,.

Under the gauge transformation \eqref{spin 2 gt} the bulk action
$\mathcal I_{\sst \mathcal M}$ is not invariant but gives a
boundary term because the gauge invariance of the linearized
Gravity action requires again an integration by part. More
precisely the gauge variation of the bulk action is given with the
aid of the Bianchi identity by the following boundary term: \be
    \delta_{\varepsilon}\, \mathcal{I}_{\sst \mathcal M}=
    \tfrac{z_{B}^{-d}}2\,
    \ppd{v\,\varepsilon}{
    \left(1-\tfrac14\,U^{2}\,\partial_{U}^{2} \right)
     \mathcal{F}\,\varphi}_{z_{B}}.
     \label{bulk gv}
\ee Note that the bulk action is still invariant under the gauge
transformation whose gauge parameter vanishes on the boundary:
$\varepsilon(z_{B})=0$\,.
We want to complete $\mathcal
I_{\sst \partial \mathcal M}^{\sst\,0}$ with suitable
$\partial_{z}\,\phi^{\sst (r)}$-free boundary terms so that the
resulting boundary term $\mathcal I_{\sst \partial \mathcal M}$
satisfies \be
    \delta_{\epsilon^{\sst (r)}}\left(\,\mathcal I_{\sst \mathcal M}
    +\mathcal I_{\sst \partial \mathcal M}\,
    \right)=0\qquad [r=1,0]\,.
    \label{tot g inv}
\ee
To begin with, we construct a boundary term which
restores the gauge invariance under  $\epsilon^{\sst (1)}$
(that is, both of $\epsilon^{\sst (1)}(z_{B})$ and
$\partial_{z}\,\epsilon^{\sst (1)}(z_{B})$).

For that, we need to first compute the gauge variations of
$\mathcal I_{\sst \mathcal M}$ and $\mathcal I^{\sst \,0}_{\sst
\partial \mathcal M}$ under $\epsilon^{\sst (1)}$\,, and then
construct a suitable boundary term to cancel them. It is in fact
more convenient to consider, instead of $\mathcal I^{\sst
\,0}_{\sst \partial \mathcal M}$\,, a complemented one $\mathcal
I_{\sst \partial \mathcal M}^{\sst\,0'}$ whose gauge variation is
easier to compute. This is equivalent since in any case
we should add another boundary term to compensate the gauge
variation. Let us replace $z\,\partial_{z}\,\phi^{\sst (2)}$
in the LHS of $\mathcal I^{\sst \,0}_{\sst \partial \mathcal M}$
\eqref{var 2} with a $\epsilon^{\sst (1)}(z_B)$,
$\partial_z\epsilon^{\sst (1)}(z_B)$ and
$\partial_z\epsilon^{\sst (0)}(z_B)$-invariant:
\be
    \chi^{\sst (2)} :=
    z\,\partial_{z}\,\phi^{\sst (2)}
    -z\,(u\cdot\partial )\,\phi^{\sst (1)}
    +\tfrac12\,u^{2}\,\phi^{\sst(0)}\,.
\ee
 The complemented boundary term containing
$\partial_{z}\,\phi^{\sst (2)}$ becomes now \be
    \mathcal{I}^{\sst\,0'}_{\sst \partial \mathcal M}=
     \tfrac{z_{B}^{-d}}2\,
    \ppd{ \phi^{\sst (2)}}{
    \left(-1+\tfrac 12\,u^{2}\,\partial_{u}^{2}\right)
    \chi^{\sst (2)}}_{z_{B}},
    \label{s2 cmp}
\ee and its $\epsilon^{\sst (1)}$-variation is given by \be
    \label{bd gv}
    \delta_{\epsilon^{\sst (1)}}\,\mathcal{I}^{\sst\,0'}_{\sst \partial \mathcal M}
    =\tfrac{z_{B}^{-d+1}}2\,
    \ppd{\epsilon^{\sst (1)}}{
    \left(\partial_{u}\cdot\partial -u\cdot\partial \,\partial_{u}^{2}\right)
    \chi^{\sst (2)}}_{z_{B}}.
\ee After obtaining the gauge variation of the boundary term, it
is time to compute the gauge variation of the bulk  action
$\mathcal I_{\sst \mathcal M}$\,. For the case of
$\epsilon^{\sst (1)}$-gauge variation, \eqref{bulk gv}
reduces  to \be
    \delta_{\epsilon^{\sst (1)}}\,\mathcal{I}_{\sst \mathcal M}=
    \tfrac{z_{B}^{-d}}2\,
    \ppd{\epsilon^{\sst (1)}}{
    \partial_{v}\,\mathcal{F}\,\varphi}_{z_{B}}\,,
    \label{2 bd variat.}
\ee and we obtain $\left(\partial_{v}\,\mathcal{F}\,\varphi\right)\!(x,z;u,0)$
 from a direct computation \eqref{F s-1} as
\ba
    && \left(\partial_{v}\,\mathcal{F}\,\varphi\right)\!(x,z;u,0) =
    z^{2}\, ( u\cdot \partial \,\partial_{u}^{2}-\partial \cdot\partial_{u})\,
    \partial_{z}\,\phi^{\sst (2)}(x,z;u) \nn
    &&\qquad +\, z^{2}\left(\Box-u\cdot\partial \,
    \partial \cdot \partial_{u} \right) \phi^{\sst (1)}(x,z;u)+
    (d-1)\,z\,u\cdot\partial \,\phi^{\sst (0)}(x,z)\,.
\ea Finally, by summing \eqref{bd gv} and \eqref{2 bd variat.}, we
see that the $\epsilon^{\sst (1)}$-variation of the total action vanish:
\be
    \delta_{\epsilon^{\sst (1)}}\left(\,\mathcal{I}_{\sst \mathcal M}+
    \mathcal{I}^{\sst\, 0'}_{\sst \partial \mathcal M}\,\right)
    =0\,.
\ee
Moreover the total action is also invariant under the gauge
transformation generated by $\partial_{z}\,\epsilon^{\sst
(0)}(z_{B})$ since $\phi^{\sst (2)}$ and $\chi^{\sst (2)}$ are invariant.
In fact, the boundary term that we have just constructed corresponds
to the quadratic part of the boundary term appearing in the linearized AdS Gravity,
that is the Einstein-Hilbert bulk action supplemented
with the York-Gibbons-Hawking term.
More precisely, $\mathcal I^{\,\sst 0'}_{\sst\partial \mathcal M}$
coincides to the boundary term \eqref{bd spin2} of the linearized AdS Gravity (see Appendix \ref{sec:lin GR} for the details).

Let us now determine the variation of the bulk and boundary
terms under the gauge transformation generated by
$\epsilon^{\sst(0)}$, the first is given by
\ba
    \delta_{\epsilon^{\sst (0)}}\,
    \mathcal{I}_{\sst \mathcal M}&=&
    \tfrac{z_B^{-d}}4\,\ppd{\epsilon^{\sst(0)}}
    {(\partial_v{}^2-\partial_u{}^2)\,\mathcal{F}\,\varphi^{\sst(2)}}\nn
    &=& \tfrac{z_B^{-d}}2\,
    \ppd{\epsilon^{\sst(0)}}{(d-1)\,\partial_u^{\,2}\,\chi^{\sst (2)}
    -z^2 \left[\Box\,\partial_u^{\,2}-(\partial_u\cdot\partial )^2\right]
    \phi^{\sst(2)}}_{z_B},
\ea
and the second
\be
    \delta_{\epsilon^{\sst (0)}}\,\mathcal{I}_{\sst \partial\mathcal M}^{\sst \,0'}
    =\tfrac{z_B^{-d}}2\,\ppd{\epsilon^{\sst (0)}}
    {-(d-1)\,\partial_u^{\,2}\,\chi^{\sst (2)}
    -z^2\left[\Box\,\partial_u^{\,2}-
    (\partial_u\cdot\partial)^2\right]\phi^{\sst(2)}}_{z_B}.
\ee
The $\chi^{\sst (2)}$-contribution cancels and finally we get for the
total variation
\be
    \delta_{\epsilon^{\sst(0)}}
    \left(\mathcal{I}_{\sst \mathcal M}+
    \mathcal{I}^{\sst\, 0'}_{\sst \partial \mathcal M}\right)
    ={-z_B^{-d+2}}\,\ppd{\epsilon^{\sst(0)}}
    {\left[\Box\,\partial_u^{\,2}-(\partial_u\cdot\partial )^2\right]
    \phi^{\sst(2)}}_{z_B}.\label{nin}
\ee
It is natural now to continue the procedure and look for a
boundary term which is $\partial_z\,\phi^{\sst (r)}$-independent,
invariant under the $\epsilon^{\sst (1)}$\,,
$\partial_z\,\epsilon^{\sst (1)}$ and
$\partial_z\,\epsilon^{\sst (0)}$ and which cancels the above
variation.
In fact, checking the gauge variations we easily see
that such a term could depend only on $\phi^{\sst (2)}$\,. The
$\epsilon^{\sst (1)}$-invariant terms  with two derivatives reduce to
\be
    \mathcal I^{\sst\rm\, (c.t.)}_{\sst \partial \mathcal M}=
    c\,z_{B}^{-d+2}\,\ppd{\phi^{\sst (2)}}{(1-\tfrac14\,{u^2}\,\partial_u^2)\,
    {\cal F}_{M^{d}}\,\phi^{\sst (2)}}_{z_{B}}\,,
\label{c.t. 2}
\ee
with an undetermined constant $c$\,.
Its variation under $\epsilon^{\sst (0)}$ is
\be
    \delta_{\epsilon^{\sst (0)}}\,\mathcal I^{\sst\rm\, (c.t.)}_{\sst \partial \mathcal M}
    =2(d-2)\,c\,z_{B}^{-d+2}\,\ppd{\epsilon^{\sst(0)}}
    {\left[\Box\,\partial_u^2-(\partial_u\cdot\partial)^2\right]
    \phi^{\sst (2)}}_{z_{B}}.
\ee
As a result, for $d\neq 2$ it is possible to cancel the term
\eqref{nin} by the addition of $\mathcal I^{\sst\rm\, (c.t.)}_{\sst \partial \mathcal M}$ with
\be
    c=\tfrac{1}{2(d-2)}\,.
\ee
For $d=2$ it is not
possible to cancel \eqref{nin} by a local term, this results in an
anomaly under the symmetries generated by $\epsilon^{\sst(0)}$\,.

In brief, we have found that in the spin 2 case the bulk action
has to be amended by a boundary term in such a way that the
equations of motion are correctly reproduced and that the action
is invariant under all the gauge symmetries but the ones generated
by $\epsilon^{\sst (0)}$ which are anomalous for $d=2$\,.
We now turn to the generalization of the above
results to the arbitrary spin case.

\subsection{General case}

We want to construct a boundary term $\mathcal{I}_{\sst
\partial \mathcal M}$ satisfying \eqref{2 bd}. In fact, it will be
more convenient as for spin 2 to first modify $\mathcal{I}^{\sst\, 0}_{\sst
\partial \mathcal M}$ into $\mathcal{I}^{\sst\, 0'}_{\sst \partial
\mathcal M}$ such that its $\epsilon^{\sst(s-1)}$-variation has a simpler expression.
Then, we construct
a boundary term $\mathcal{I}^{\sst\, 1}_{\sst \partial \mathcal
M}$  such that
\be
    \delta_{\alpha}\,\mathcal{I}^{\sst\,1}_{\sst \partial \mathcal M}=-\,
    \delta_{\alpha}\left(\,
    \mathcal{I}_{\sst \mathcal M} + \mathcal{I}^{\sst\,0'}_{\sst \partial \mathcal M}\,\right),
    \label{fst bdc}
\ee
where $\alpha$ is a gauge transformation generated by $\epsilon^{\sst(s-1)}$
and $\partial_z\epsilon^{\sst(s-1)}$\,.
We will proceed in steps, from $\epsilon^{\sst (s-1)}$ and $\partial_{z}\epsilon^{\sst (s-1)}$
transformations we will construct the first boundary term
$\mathcal{I}^{\sst\, 1}_{\sst \partial \mathcal M}$ and then
from the transformation of $\mathcal{I}_{\sst \mathcal M} +
\mathcal{I}^{\sst\,0'}_{\sst \partial \mathcal M}+
\mathcal{I}^{\sst\, 1}_{\sst \partial \mathcal M}$ under $\partial_z \epsilon^{\sst(s-2)}$
we determine
$\mathcal{I}^{\sst\, 2}_{\sst \partial \mathcal M}$\,.

The zeroth boundary term $\mathcal{I}^{\sst\,0}_{\sst \partial
\mathcal M}$ \eqref{bd. term exp} contains
$\partial_{z}\,\phi^{\sst (s)}$ and
$\partial_{z}\,\psi^{\sst (s-3)}$ on the RHS of the brackets.
They can be replaced by the  invariants under the
$\epsilon^{\sst (s-1)}$-transformation by adding
$\partial_{z}\,\phi^{\sst (r)}$-independent terms as \ba
    \chi^{\sst (s)} &\!:=\!&
    (z\,\partial_{z}+s-2)\,\phi^{\sst (s)}
    -z\,(u\cdot\partial )\,\phi^{\sst (s-1)}-u^{2}\,\phi^{\sst(s-2)}\,,\\
    \chi^{\sst (s-3)} &\!:=\!&
    (z\,\partial_{z}+s-3)\,\psi^{\sst (s-3)}
    + (d+2s-5)\,\phi^{\sst (s-3)} +\tfrac12\,u^{2}\,\partial^{2}_{u}
    \left(\phi^{\sst (s-3)}+\partial_{u}^{2}\,\phi^{\sst (s-1)}\right), \nonumber
\ea so that  $\mathcal{I}^{\sst\,0}_{\sst
\partial \mathcal M}$ is replaced by \ba
    \mathcal{I}^{\sst\, 0'}_{\sst \partial \mathcal M}=
     \tfrac{z_{B}^{-d}}2\,\Big[
    \ppd{ \phi^{\sst (s)}}{\mathsf{P}_{\rm e}'\,\chi^{\sst (s)}}_{z_{B}}
    +
    \ppd{ \psi^{\sst (s-3)}}{
    \mathsf{P}_{\rm o}'\,\chi^{\sst (s-3)}}_{z_{B}} \Big]\,.
    \label{new bd. term exp}
\ea Since $\chi^{\sst (s)}$ and $\chi^{\sst (s-3)}$ are
$\epsilon^{\sst (s-1)}$-invariants, the $\epsilon^{\sst(s-1)}$-variation
of the above is simple to compute, and is
given  by \be
    \delta_{\epsilon^{\sst (s-1)}}\,\mathcal{I}^{\sst\,0'}_{\sst \partial \mathcal M}
    =\tfrac{z_{B}^{-d}}2\,
    \ppd{\epsilon^{\sst (s-1)}}{
    \mathsf{P}_{\rm e}\left[z\,(\partial_{u}\cdot\partial -
    u\cdot\partial \,\partial_{u}^{2})
    \,\chi^{\sst (s)}
    +u^{2}\,\chi^{\sst (s-3)}\right]}_{z_{B}}\,,
    \label{variat.}
\ee with \be
    \mathsf{P}_{\rm e}:=\sum_{n=0}^{\infty}
    \tfrac{1}{(2n)!}\,(u^{2})^{n} (\partial_{u}^{2})^{n}\,,
    \qquad
    \mathsf{P}_{\rm o}:=\sum_{n=0}^{\infty}
    \tfrac{1}{(2n+1)!}\,(u^{2})^{n} (\partial_{u}^{2})^{n}\,.
\ee In order to get \eqref{variat.}, we have used the identities (\ref{P id1}\,,\,\ref{P id2}).
To construct the first boundary term $\mathcal I_{\sst \partial
\mathcal M}^{\sst\, 1}$ with \eqref{fst bdc}, we need to compute
also the gauge variation of the bulk action. From the gauge
invariance of the Fronsdal equation and the Bianchi identity
\eqref{Bianchi}, we get the gauge variation of the action as \eqref{bulk gv}.
 For a more explicit expression of the above, we consider the
radial decomposition of the Fronsdal equation: \be
    (\mathcal{F}\,\varphi^{\sst (s)})(x,z;u,v) = \sum_{r=0}^{s}\,\frac{v^{r}}{r!}\,
    f^{\sst(s-r)}(x,z;u)\,,
    \label{Frons dec}
\ee where $f^{\sst (r)}$ are determined by $f^{\sst (s)},
f^{\sst (s-1)}, f^{\sst (s-2)}$ and $f^{\sst (s-3)}$ since
the Fronsdal equation also has vanishing the double-trace: \ba
    f^{\sst (s-2n-m)}(x,z;u)  = (-1)^{n+1}
    &&\hspace{-5pt} \Big[\, (n-1)\,(\partial_{u}^{2})^{n} \left(
    \partial_{v}^{m}\,\mathcal{F}\,\varphi
    \right)\!(x,z;u,0)\nn
    &&  +\, n\,(\partial_{u}^{2})^{n-1} \left(\partial_{v}^{m+2}\,
    \mathcal{F}\,\varphi\right)\!(x,z;u,0)\, \Big]\,,
    \label{Frons dtc}
\ea where $m$ is either 0 or 1. Using \eqref{Frons dec} and
\eqref{Frons dtc}, the gauge variation of the bulk action is
written as \ba
    \delta_{\varepsilon}\,\mathcal{I}_{\sst \mathcal M} \e
    \tfrac{z_{B}^{-d}}2\,\Big[
    \ppd{\epsilon^{\sst (s-1)}}{
    \mathsf{P}_{\rm e}\,\partial_{v}\,\mathcal{F}\,\varphi}
    + \ppd{\epsilon^{\sst (s-2)}}{\mathsf{P}_{\rm o}\,
    \tfrac12\,(\partial_{v}^{2}-\partial_{u}^{2})\,\mathcal{F}\,\varphi}\Big]\,.
    \label{bd variat.}
\ea The expressions for $\partial_{v}\,\mathcal{F}\,\varphi\,|_{v=0}$ and $
(\partial_{u}^{2}-\partial_{v}^{2})\,\mathcal{F}\,\varphi\,|_{v=0}$
are obtained in \eqref{F s-1} and \eqref{F s-2}
from the explicit decomposition \eqref{Fronsdal expl.} of the
Fronsdal operator.  Noting that
$\partial_{v}\,\mathcal{F}\,\varphi\,|_{v=0}$ is gauge
invariant as the Fronsdal equation is, we express it in terms of
$\epsilon^{\sst (s-1)}$-invariants as \be
\partial_{v}\,\mathcal{F}\,\varphi\,\big|_{v=0}=
   -z\,(\partial_{u}\cdot\partial -u\cdot\partial \,\partial_{u}^{2})
    \,\chi^{\sst (s)}
    -u^{2}\,\chi^{\sst (s-3)}
    +z\,u\cdot\partial \,\zeta^{\sst (s-2)}\,,
\ee where $\zeta^{\sst (s-2)}$ is a $\epsilon^{\sst(s-1)}$-invariant: \be
    \zeta^{\sst (s-2)} :=
    z\,u\cdot\partial \,\psi^{\sst (s-3)}
    +3\,(d+2s-5)\,\phi^{\sst (s-2)}+u^{2}\,\partial_{u}^{2}
    \left(\phi^{\sst (s-2)}+2\,\partial_{u}^{2}\,\phi^{\sst (s)}\right).
\ee

Finally, summing up the two contributions (\ref{variat.}\,,\,\ref{bd variat.}),
 the $\epsilon^{\sst (s-1)}$-variation
of the bulk action plus the zeroth boundary term reads \be
    \delta_{\epsilon^{\sst (s-1)}}
    \Big(\,\mathcal{I}_{\sst \mathcal M}
    +\mathcal{I}^{\sst\, 0'}_{\sst \partial \mathcal M}\,\Big)
    =\tfrac{z_{B}^{-d}}2\,
    \ppd{\epsilon^{\sst (s-1)}}
    {\mathsf{P}_{\rm e}\,z\,{u}\cdot\partial \,\zeta^{\sst
(s,s-2)}}_{z_{B}}\,.
    \label{var s}
\ee
We now look for a local $\partial_z\,\phi^{\sst (r)}$-independent expression
whose variation is precisely \eqref{var s}.

Using an identity of $\mathsf{P}$-operator \eqref{P id2}
and an integration by parts, the variation \eqref{var s} becomes
    \be
 -\tfrac{z_{B}^{-d}}2\,
    \ppd{
z\left(\partial_{u}\cdot\partial +u\cdot\partial
\,\partial_{u}^{2}\right)
    \epsilon^{\sst (s-1)}\,}
    {\mathsf{P}_{\rm o}\,\zeta^{\sst(s-2)}}_{z_{B}}\,. \ee So it can be cancelled by a term of the
form \be
    \mathcal{I}_{\sst \partial \mathcal M}^{\sst\,1}=
    \tfrac{z_{B}^{-d}}2\,
    \ppd{\xi^{\sst (s-2)}}{\mathsf{P}_{\rm o} \,\zeta^{\sst (s-2)}}_{z_{B}},
\ee where $\xi^{\sst (s-2)}$ is a linear combination of the
fields which does not involve any $\partial_{z}\,\phi^{\sst (r)}$
and whose  $\epsilon^{\sst (s-1)}$-variation  is
\be
    \delta_{\epsilon^{\sst (s-1)}}\,\xi^{\sst (s-2)}
    =z\left(\partial_{u}\cdot\partial +u\cdot\partial \,\partial_{u}^{2}\right)
    \epsilon^{\sst (s-1)}\,.
\ee
From the gauge transformations \eqref{decomp. bd necessaries}
of the decomposed fields, we determine  $\xi^{\sst (s-2)}$ as
\be
    \xi^{\sst (s-2)}=\tfrac12\, \left(\partial_{u}^{2}\,\phi^{\sst (s)}
    -\phi^{\sst (s-2)}\right)\,.
    \label{xi}
\ee

Now we turn to the determination of  the second
($\epsilon^{\sst (s-1)}$-invariant) boundary term $\mathcal
I^{\sst \,2}_{\sst
\partial \mathcal M}$ by requiring \eqref{2 bd} under
the $\partial_{z}\epsilon^{\sst (s-2)}$-variation. For that, let us
first consider the vanishing boundary condition for the gauge
parameter: $\epsilon^{\sst (s-2)}(z_{B})=0$, but with an
arbitrary $\partial_{z}\,\epsilon^{\sst (s-2)}(z_{B})$\,.
After finding $\mathcal I^{\sst
\,2}_{\sst
\partial \mathcal M}$ with this condition, we will see, in the
next subsection, whether this boundary condition can be relaxed by
adding another boundary term.

Since $\delta_{\epsilon^{\sst (s-2)}}\,\mathcal I_{\sst
\mathcal M}=0$ with $\epsilon^{\sst (s-2)}(z_{B})=0$\,, we
require simply \be
     \delta_{\epsilon^{\sst (s-2)}}\,
     \mathcal{I}^{\sst\,2}_{\sst \partial \mathcal M}=-\,
    \delta_{\epsilon^{\sst (s-2)}}\left(\,
    \mathcal{I}^{\sst\,0'}_{\sst \partial \mathcal M}
    +\mathcal{I}^{\sst\,1}_{\sst \partial \mathcal M}\,\right),
    \label{11 inv}
\ee where the zeroth and the first boundary terms are obtained in
the previous section as \be
    \mathcal{I}_{\sst \partial \mathcal M}^{\sst\,0'}
     +\mathcal{I}_{\sst \partial \mathcal M}^{\sst\,1}=  \tfrac{z_{B}^{-d}}2\,
     \Big[
    \ppd{ \phi^{\sst (s)}}{\mathsf{P}_{\rm e}'\,\chi^{\sst (s)}}_{z_{B}}
    +
    \ppd{ \psi^{\sst (s-3)}}{
    \mathsf{P}_{\rm o}'\,\chi^{\sst (s-3)}}_{z_{B}}+\ppd{\xi^{\sst (s-2)}}{
    \mathsf{P}_{\rm o}\,\zeta^{\sst (s-2)}}_{z_{B}}\Big]\,.
\ee The above is written in terms of the boundary values of
$\phi^{\sst (s)},\ \chi^{\sst (s)},\ \psi^{\sst (s-3)},\
\chi^{\sst (s-3)},\ \xi^{\sst (s-2)}$ and $\zeta^{\sst
(s,s-2)}$\,, and their boundary values transform under
$\epsilon^{\sst (s-2)}$ with $\epsilon^{\sst(s-2)}(z_{B})=0$ as \ba
    &\delta\, \phi^{\sst (s)}=0\,,\qquad
    & \delta\,\chi^{\sst (s)}=
    -3\,u^{2}\,z\,\partial_{z}\,\epsilon^{\sst (s-2)}\,, \nn
    &\delta\, \psi^{\sst (s-3)}=0\,,\qquad
    & \delta\,\chi^{\sst (s-3)}=
    z\left(3\,\partial \cdot\partial_{u}
    +u\cdot\partial \,\partial_{u}^{2}\right)
    \partial_{z}\,\epsilon^{\sst (s-2)}\,, \\
    & \delta\, \xi^{\sst (s-2)}=
    -z\,\partial_{z} \,\epsilon^{\sst (s-2)}\,,\qquad
    & \delta\,\zeta^{\sst (s-2)}=
    2\left[3(d+2s-5)+\,u^{2}\,\partial_{u}^{2}\right]
    z\,\partial_{z}\,\epsilon^{\sst (s-2)}\,.\nonumber
\ea
By using the
above transformations rules, we get the $\epsilon^{\sst(s-2)}$-variation of the boundary term,
with $\epsilon^{\sst (s-2)}(z_{B})=0$\,, as
\ba
    \delta_{\epsilon^{\sst (s-2)}}
    \left(\, \mathcal{I}_{\sst \partial \mathcal M}^{\sst\,0'}
    +\mathcal{I}_{\sst \partial \mathcal M}^{\sst\,1}\,\right)=
    -z_{B}^{-d}\,
    \ppd{ \mathsf{P}_{\rm o}\,z\,\partial_{z}\, \epsilon^{\sst (s-2)}}
    {\zeta^{\sst (s-2)}}_{z_{B}}.
    \label{s3 v2}
\ea
This variation can be compensated by the following second
boundary term:
\be
    \mathcal I^{\sst\,2}_{\sst \partial \mathcal M}=
    \tfrac{z_{B}^{-d}}{4}\,
    \ppd{\zeta^{\sst (s-2)}}{\mathsf P_{\rm o}''\,\zeta^{\sst (s-2)}}_{z_{B}},
    \qquad
    \mathsf P_{\rm o}'' := \mathsf P_{\rm o}\,\left[3\,(d+2s-5)+u^{2}\,\partial_{u}^{2}\right]^{-1}\,.
    \label{zeta zeta}
\ee
The inversion of the operator can be readily done and the result is
\be
    \mathsf T :=
    \left[3\,(d+2s-5)+u^{2}\,\partial_{u}^{2}\right]^{-1}
    = \frac1{3(d+2s-5)}
    \sum_{n=0}^{\infty}
    \frac1{4^n\,(a_+)_n\,(a_-)_n}
    (u^2)^n(\partial_u^2)^n\,,
\ee
with
\be
    a_\pm=\tfrac{d+2\,u\cdot\partial_u+2}4
    \pm \sqrt{\left(\tfrac{d+2\,u\cdot\partial_u-2}4\right)^2
    +\tfrac{3(d+2s-5)}{4}}\,.
\ee
 Notice that this second boundary term has a peculiar aspect. It
contains, for $s>2$,  a second order derivative term in $x$ (since
$\zeta^{\sst (s-2)}$ contains a term with one derivative).

It remains to examine whether the symmetries under
$\epsilon^{\sst(s-2)}(z_B)$ can be restored.
For that we need first to obtain
the variation of $\mathcal I_{\sst \mathcal M}+\mathcal I_{\sst \partial\mathcal M}$
under $\epsilon^{\sst (s-2)}$\,,
and it is given by (see Appendix \ref{sec: gauge var} for more details)
\be
    \delta_{\epsilon^{\sst (s-2)}}\left(\mathcal I_{\sst \mathcal M}+\mathcal I_{\sst \partial\mathcal M}\right)
    = \tfrac{z_{B}^{-d}}2\,\ppd{\epsilon^{\sst (s-2)}}{J_{\sst\mathcal M}+J_{\sst \partial\mathcal M}}_{z_{B}}\,,
\ee
with
\ba
    J_{\sst\mathcal M}+J_{\sst \partial\mathcal M}\e
     \mathsf P_{\rm o}\,\Big\{
    z^{2}\,\Big[\,2\left((\partial_{u}\cdot\partial)^{2}-\Box\,\partial_{u}^{2}\right)
    -\tfrac52\,u\cdot\partial\,\partial_{u}\cdot\partial\,\partial_{u}^{2}-\tfrac32\,(u\cdot\partial)^{2}\,(\partial_{u}^{2})^{2}\nn
    && \hspace{40pt}+\,\tfrac12\,u\cdot\partial\,(u\cdot\partial\,\partial_{u}^{2}+3\,\partial_{u}\cdot\partial)
    \left(1+4\,\mathsf T\,u^{2}\,\partial_{u}^{2}\right) \partial_{u}^{2}\,\Big]\,\phi^{\sst (s)}\nn
    &&\hspace{25pt} +\,\Big[ \left(s-2+\tfrac12\,u^{2}\,\partial_{u}^{2}\right)\left(3(d+2s-5)+u^{2}\,\partial_{u}^{2}\right)   \nn
    && \hspace{40pt}
    -\tfrac12 \left(7d+10s-19-u^{2}\,\partial_{u}^{2}\right) u^{2}\,\partial_{u}^{2}
    -4\,\partial_{u}^{2}\,(u^{2})^{2}\,\partial_{u}^{2}\,u^{2}\,\partial_{u}^{2}\,\mathsf T
    \Big]\,\partial_{u}^{2}\,\phi^{\sst (s)} \nn
    && \hspace{25pt} +\,z^{3}\,u\cdot\partial \left(u\cdot\partial+3\,\partial_{u}\cdot\partial\right)\mathsf T\,u\cdot\partial\,\psi^{\sst (s-3)}\nn
    && \hspace{25pt} +\,z\left[
    (d+2s-5+u^{2}\,\partial_{u}^{2})\,u\cdot\partial+u^{2}\,\partial_{u}\cdot\partial -2\,\partial_{u}^{2}\,(u^{2})^{2}\,\partial_{u}^{2}\,\mathsf T\,u\cdot\partial
    \right] \psi^{\sst (s-3)}\,\Big\}\nn
    &&\quad-\,(d+2s-5)\left[
    (3(s-2)+u^{2}\,\partial_{u}^{2})\,\mathsf P_{\rm e}\,\partial_{u}^{2}\,\phi^{\sst (s)}
    +z\,u\cdot\partial\,\mathsf P_{\rm e}\,\psi^{\sst (s-3)}\right].
\label{var ep s-2}
\ea
Notice that the variation does not contain $\phi^{\sst(s-2)}$
while the dependence on $\phi^{\sst (s-1)}$ is only
through $\psi^{\sst (s-3)}$\,.
In the next section, we show that the variation \eqref{var ep s-2}
cannot be cancelled by including additional boundary terms.

Before closing this section let us assemble all the terms and give
the complete  boundary term
\ba
    \mathcal{I}_{\sst \partial \mathcal M}
     \e \tfrac{z_{B}^{-d}}2\,
     \Big[
    \ppd{ \phi^{\sst (s)}}{\mathsf{P}_{\rm e}'\,\chi^{\sst (s)}}_{z_{B}}
    +\tfrac{1}{2}\,
    \ppd{3\,\partial_u^{\,2}\,\phi^{\sst (s-1)}+\phi^{\sst (s-3)}}{
    \mathsf{P}_{\rm o}'\,\chi^{\sst (s-3)}}_{z_{B}}\nonumber\\
    &&\qquad +\, \tfrac{1}{2}\,
    \ppd{\partial_u^{\,2}\,\phi^{\sst (s)}-\phi^{\sst (s-2)}}{
    \mathsf{P}_{\rm o}\,\zeta^{\sst (s-2)}}_{z_{B}} +
    \tfrac{1}{2}\,
    \ppd{\zeta^{\sst (s-2)}}{\mathsf P_{\rm o}''\,\zeta^{\sst
    (s,s-2)}}_{z_{B}}\Big]\,,\qquad
\label{boundary}
\ea where we recall the definitions \ba
    \chi^{\sst (s)} &\!:=&
    (z\,\partial_{z}+s-2)\,\phi^{\sst (s)}
    -z\,(u\cdot\partial )\,\varphi^{\sst
    (s,s-1)}-u^{2}\,\phi^{\sst (s-2)}\,,\nn
    \chi^{\sst (s-3)} &\!:=&
    \tfrac 12\,(z\,\partial_{z}+s-3)
     \left(3\,\partial_{u}^{\,2}\,\phi^{\sst (s-1)}+\phi^{\sst (s-3)}\right)+\nn
    &&\ +\,(d+2s-5)\,\phi^{\sst (s-3)} +\tfrac12\,u^{2}\,\partial^{2}_{u}
    \left(\phi^{\sst (s-3)}+\partial_{u}^{2}\,\phi^{\sst (s-1)}\right),
    \nonumber\\
    \zeta^{\sst (s-2)} &\!:=&
    \tfrac12\,z\,u\cdot\partial \left(3\,\partial_{u}^{\,2}\,
    \phi^{\sst (s-1)}+\phi^{\sst (s-3)}\right)+\nn
    &&\ +\,3\,(d+2s-5)\,\phi^{\sst (s-2)}+u^{2}\,\partial_{u}^{2}
    \left(\phi^{\sst (s-2)}+2\,\partial_{u}^{2}\,\phi^{\sst (s)}\right).
\label{defs}
\ea

\section{General boundary actions and anomalies}
\label{sec: gen bd}

In order to see whether it is possible to
maintain the symmetry under $\epsilon^{\sst(s-2)}(z_B)$
in the presence of a boundary we will
proceed in two steps. First, we determine the most general action
compatible with the equations of motion and the invariance under
$\epsilon^{\sst (s-1)}(z_B)$\,,
$\partial_z\,\epsilon^{\sst (s-1)}(z_B)$ and
$\partial_z\,\epsilon^{\sst (s-2)}(z_B)$
and then we see whether this action can be invariant
$\epsilon^{\sst (s-2)}(z_B)$\,.

The most general action differs from \eqref{boundary} by a
quadratic term, $\mathcal{I}^{\rm\sst\,(c.t.)}_{\sst \partial\mathcal M}$, with no
$\partial_z\,\phi^{\sst (r)}$ dependence and
invariant under the transformations generated by
$\epsilon^{\sst (s-1)}(z_B)$, $\partial_z\,\epsilon^{\sst
(s-1)}(z_B)$ and $\partial_z\,\epsilon^{\sst(s-2)}(z_B)$\,.
\begin{itemize}
\item
Invariance under $\partial_z\,\epsilon^{\sst(s-2)}(z_B)$
implies that $\mathcal{I}^{\sst\rm\,(c.t.)}_{\sst \partial\mathcal M}$ does not depend on
$\phi^{\sst (s-2)}$\,;
\item
invariance under
$\partial_z\,\epsilon^{\sst (s-1)}(z_B)$ is insured if the
dependence on $\phi^{\sst (s-1)}$ and $\phi^{\sst(s-3)}$
is only through the combination $\psi^{\sst (s-3)}$ or equivalently
\be
    \beta^{\sst (s-3)}:={\mathsf T}\,\psi^{\sst (s-3)}\,.
\ee
\end{itemize}
The latter has the advantage of a simple transformation
under $\epsilon^{\sst (s-1)}(z_B)$\,:
\be
    \delta_{\epsilon^{\sst (s-1)}}\,
    \beta^{\sst (s-3)}=\partial_u^{\,2}\,\epsilon^{\sst (s-1)}\,.
\ee
In the following, we examine all $\epsilon^{\sst (s-1)}(z_B)$-invariants
built in terms of $\phi^{\sst (s)}$ and $\beta^{\sst (s-3)}$\,.
Since the gauge variation \eqref{var ep s-2} contains up to
two derivatives acting on $\phi^{\sst (s)}$\,,
we focus on the ones involving at most two derivatives on $\phi^{\sst (s)}$\,.

\subsection{Gauge-invariant boundary terms and their transformations}

To examine all invariants,
we classify the latter as the ones quadratic or linear
in the basic invariants, which are linear in  $\phi^{\sst (s)}$ and $\beta^{\sst (s-3)}$.\footnote{This is
similar to the problem of unconstrained gauge invariance with a
compensator examined in \cite{Francia:2005bu,Francia:2007qt,Francia:2008hd}
and in particular the case $k=1$ in \cite{Francia:2007qt}.}
There are only two basic invariants up to two derivatives on $\phi^{\sst (s)}$\,,
and they are
\ba
    && A^{\sst (s)}
    :=z^{2}\,{\cal F}_{M^{d}}\, \phi^{\sst (s)}-\tfrac12\,
    z^{3}\,(u\cdot\partial)^3\,\beta^{\sst (s-3)}\,,\nn
    && C^{\sst (s-4)}:=
    (\partial_u^{\,2})^2\,\phi^{\sst (s)}
    - z\left(4\, \partial_u\cdot\partial \,
    + u\cdot\partial \,\partial_u^{\,2}\right)
    \beta^{\sst (s-3)}\,,
    \label{def A C}
\ea
where $\mathcal F_{M^{d}}$ is the $d$-dimensional Fronsdal operator:
\be
    \mathcal F_{M^{d}}:=\Box - u\cdot\partial\,\partial_{u}\!\cdot\partial
    +\tfrac12\,(u\cdot\partial)^{2}\,\partial_{u}^{\,2}\,.
\ee
Notice that these two invariants are related as
\ba
\label{rel1 A C}
    &(\partial_{u}^{\,2})^{2}\, A^{\sst (s)}
    =3\,z^{2} \left[\,\Box+u\cdot\partial\,\partial_{u}\!\cdot\partial
    +\tfrac16\,(u\cdot\partial)^{2}\,\partial_{u}^{\,2}\,\right] C^{\sst (s-4)}\,,\\
\label{rel2 A C}
    &z\left(\partial_u\cdot\partial-\tfrac{1}{2}\,u\cdot\partial \,\partial_u^2
    \right){A}^{\sst (s)}+\tfrac{1}{4}\,z^{3}\,(u\cdot\partial )^3\,{C}^{\sst (s-4)}=0\,.
\ea
Hence, the double trace and the de Donder divergence of $A^{\sst (s)}$ are given by
$C^{\sst (s-4)}$\,.

Concerning the quadratic terms built with these two invariants,
there is only one invariant involving up to two derivatives\,:
\be
    \mathcal{I}^{\,\sst 1\,{\rm(c.t.)}}_{\sst \partial\mathcal M}
    =z_{B}^{-d}\,\ppd{C^{\sst (s-4)}}{\mathcal O\ C^{\sst (s-4)}}_{z_{B}}\,,
    \label{c}
\ee
where $\mathcal O$ is
a self-adjoint operator in $u$ and $x$ at most second-order in $z\,\partial$\,.
The quadratic terms in $A^{\sst (s)}$ contain at least four derivatives and
the bilinear (cross) terms in $A^{\sst (s)}$ and $C^{\sst (s-4)}$ are either
containing higher derivatives or re-expressible as \eqref{c} using \eqref{rel1 A C}.
The explicit form $\mathcal O$ is to be determined
by requiring the invariance of the total action under $\epsilon^{\sst (s-2)}$\,.

The linear invariant in $A^{\sst (s)}$ and $C^{\sst (s-4)}$ is given by
\be
    \mathcal I^{\sst\,2\,{\rm(c.t.)}}_{\sst\partial\mathcal M}
    =z_{B}^{-d}\,\Big[\
    \ppd{\phi^{\sst (s)}}{G^{\sst (s)}}_{z_{B}}\!+\,
    \ppd{\beta^{\sst(s-3)}}{K^{\sst (s-3)}}_{z_{B}} \Big]\,,
\label{c.t. 2}
\ee
where ${G}^{\sst (s)}$ and ${K}^{\sst (s-3)}$ are
linear invariants so that, with the restriction of number of derivatives,
they are given by $A^{\sst (s)}$ and $C^{\sst (s-4)}$\,.
The gauge invariance of  $\mathcal I^{\sst\,2\,{\rm(c.t.)}}_{\sst\partial\mathcal M}$
under $\epsilon^{\sst (s-1)}$
imposes the Bianchi
identities on these objects:\footnote{
Recall that the gauge transformation of
$\phi^{\sst (s)}$ is $z\,u\cdot\partial\,\epsilon^{\sst
(s-1)}$\,, this explains the $z$ factor in (\ref{inv2}).}
 \be
    z\,\partial_{u}\cdot\partial \,{G}^{\sst (s)}=u^2\,{K}^{\sst (s-3)}\,.
    \label{inv2}
\ee
Using the identity \eqref{rel2 A C}\,, we find
the solution for \eqref{inv2} as
\ba
&& {G}^{\sst (s)} =
c\,\big[ \left(1-\tfrac14\,u^2\,\partial_u^2 \right)
{A}^{\sst (s)} +\tfrac18\,
z^{2}\,(u\cdot\partial)^2\,u^2\,{C}^{\sst (s-4)}\,\big]\,,\nn
&& {K}^{\sst (s-3)} =-c\,
\tfrac z4\,\partial\cdot\partial_u\left[\,\partial_u^2\,
{A}^{\sst (s)}-\tfrac12\,z^{2}\,(u\cdot\partial)^2\,{C}^{\sst (s-4)}\,\right],
\label{GK}
\ea
where $c$ is an arbitrary constant.

We now calculate the variation of the total counterterm,
$\mathcal I^{\sst\,{\rm(c.t.)}}_{\sst\partial\mathcal M}=
\mathcal I^{\sst\,1\,{\rm(c.t.)}}_{\sst\partial\mathcal M}+
\mathcal I^{\sst\,2\,{\rm(c.t.)}}_{\sst\partial\mathcal M}$\,,
under the remaining gauge transformation, that is, the one by $\epsilon^{\sst (s-2)}(z_{B})$\,.
We obtain the variation of the counterterms as
\be
 \delta_{\epsilon^{\sst (s-2)}}\,
\mathcal I^{\sst\,{\rm(c.t.)}}_{\sst\partial\mathcal M}=
\tfrac{z_{B}^{-d}}2\,\ppd{\epsilon^{\sst (s-2)}}
{J^{\sst\,{\rm(c.t.)}}_{\sst\partial\mathcal M}}_{z_{B}}\,,
\ee
with
\ba
J^{\sst\,{\rm(c.t.)}}_{\sst\partial\mathcal M} \e
-4\,\Big[\,4(d+2s-6)\,u^2+(u^2)^2\,\partial_u^2+ \nn
&& \qquad+\,
z^2\,(3\,u\cdot\partial+u^2\,\partial_u\cdot\partial)\,\mathsf T
(4\,u\cdot\partial+u^2\,\partial_u\cdot\partial)\,\Big]\,
{\mathcal O}\,C^{\sst (s-4)}\,,\nn
&&+\,
c\,\Big\{
2(d+2s-6)\,\partial_u^2\,{A}^{\sst (s)}
-z^{2}\left[(d+2s-4)\,(u\cdot\partial)^2+
u^{2}\left(2\,\Box+u\cdot\partial\,\partial_u\cdot\partial\right)\right]
{C}^{\sst (s-4)}\nn
&&\qquad-\,z^2\,(u^2\,\partial_u\cdot\partial+3\,u\cdot\partial)\,\mathsf T\,
\partial_u\cdot\partial
\left[\,\partial_u^2\,{A}^{\sst (s)}-\tfrac{z^{2}}2\,(u\cdot\partial)^2\,{C^{\sst (s-4)}}\right]
\Big\}\,.
\ea
For more details on the calculation, see Appendix \ref{sec: counter}.
An important property of the variations of the invariants that
we obtained is that the variation involves terms with different
number of derivatives. The origin of this property lies in the
transformation of $\phi^{\sst (s)}$ and $\beta^{\sst (s-3)}$ under $\epsilon^{\sst (s-2)}$
with respectively zero and one derivative.

\subsection{Anomaly}
\label{subsec: Anomaly}

Now we determine the variation under $\epsilon^{\sst (s-2)}$ of the total action:
\be
    \delta_{\epsilon^{\sst (s-2)}}\big(\,
    \mathcal I_{\sst \cal M}+\mathcal I_{\sst \partial \cal M}+
    \mathcal I^{\sst {\,\rm (c.t.)}}_{\sst \partial \cal M}\,\big)
    =\ppd{\epsilon^{\sst (s-2)}}{\mathscr A^{\sst (s-2)}}_{z_{B}}\,,
\ee
 and see whether there exists a choice for the counterterms such that this variation vanishes.
 In other words, we examine whether
 there exist an operator ${\cal O}$ \eqref{c} and constants $c$  such that the anomaly:
 \be
    \mathscr A^{\sst (s-2)}:=
    \tfrac{z^{-d}}2\left( J_{\sst \mathcal M}+J_{\sst \partial \mathcal M}
    +J_{\sst \mathcal \partial \mathcal M}^{\sst \,{\rm (c.t.)}}
    \right),
\ee
vanishes. In the last section, we have shown that
$J_{\sst \mathcal \partial \mathcal M}^{\sst \,{\rm (c.t.)}}$ is given
in terms of $A^{\sst (s)}$ and $C^{\sst (s-4)}$\,.
Actually, the gauge variation $J_{\sst \mathcal M}+J_{\sst \partial \mathcal M}$
\eqref{var ep s-2}
can also be expressed in terms of these invariants as
\ba
    J_{\sst\mathcal M}+J_{\sst \partial\mathcal M}\e
    -\,\mathsf P_{\rm o}\,\partial_{u}^{2}\,A^{\sst (s)}
    -\,\mathsf P_{\rm o}\,\tfrac{z^{2}}2 \left[(u\cdot\partial)^{2}-4\,u\cdot\partial
    \left(u\cdot\partial\,\partial_u^2+3\,\partial_{u}\cdot\partial\right)\mathsf T\,u^{2}\right] C^{\sst (s-4)}\nn
    &&-\,\Big\{\,
    \mathsf P_{\rm o}\left[s+2d-\tfrac12\,(3d+8s-18)\,u^{2}\,\partial_{u}^{2}+4\,\partial_{u}^{2}\,(u^{2})^{2}\,\partial_{u}^{2}\,\mathsf T\right] u^{2}\nn
    && \qquad
    +\,(d+2s-5)\left(\mathsf P_{\rm e}\,u^{2}+3\,(s-2)\,u^{2}\,\mathsf P_{\rm o}'\right)\Big\}\,C^{\sst (s-4)}\,.
\label{anomal}
\ea
In order to see whether the anomaly can be canceled,
let us first focus on the part of $\mathscr A^{\sst (s-2)}$ proportional to
the trace of $A^{\sst (s)}$\,,
then we get
\be
    z^{-d}\left[-\tfrac12
    +c \left(d+2s-6-z^2\,
    \tfrac{(u^2\,\partial_u\!\cdot\,\partial+3\,u\cdot\partial)\,
    \partial_u\cdot\partial}{6(d+2s-5)}\,
    \right)\right] \partial_u^2\,{A}^{\sst (s)}\,.
\ee
For $s>2$ where $\partial_u\cdot\partial\partial_u^2\,{A}^{\sst
(s)}$ is not zero this term cannot be canceled by appropriately choosing $c$\,,
and it is independent of $C^{\sst (s-4)}$\,.
Therefore, there is no boundary counterterm which
leads to the cancellation of this term implying that
the symmetry under $\epsilon^{\sst (s-2)}$ cannot be restored.

Although the counterterm cannot restore the gauge symmetries,
it can modify the anomaly $\mathscr A^{\sst (s-2)}$\,.
By focusing now on the part of $\mathscr A^{\sst (s-2)}$ proportional to $C^{\sst (s-4)}$\,,
we get
\ba
    && -\,z^{-d+2}\,\Big\{\,\mathsf P_{\rm o}\left[\,
    \tfrac14(u\cdot\partial)^{2}-u\cdot\partial
    \left(u\cdot\partial\,\partial_u^2+3\,\partial_{u}\cdot\partial\right)\mathsf T\,u^{2}\,\right]\nn
    &&\hspace{50pt}
    +\,2\,(3\,u\cdot\partial+u^2\,\partial_u\cdot\partial)\,\mathsf T
    (4\,u\cdot\partial+u^2\,\partial_u\cdot\partial)\,
    {\mathcal O}\,\Big\}\,C^{\sst (s-4)}\nn
    &&-\,\tfrac{z^{-d}}2\,\Big\{\,
    \mathsf P_{\rm o}\left[s+2d-\tfrac12\,(3d+8s-18)\,u^{2}\,\partial_{u}^{2}+4\,\partial_{u}^{2}\,
    (u^{2})^{2}\,\partial_{u}^{2}\,\mathsf T\right] u^{2} \\
    && \hspace{35pt}
    +\,(d+2s-5)\left(\mathsf P_{\rm e}\,u^{2}+3\,(s-2)\,u^{2}\,\mathsf P_{\rm o}'\right)
    + 4 \left[4(d+2s-6)\,u^2+(u^2)^2\,\partial_u^2\right] {\mathcal O}\,\Big\}\,C^{\sst (s-4)}\,,
    \nonumber
\ea
where we have chosen $c=0$\,.
Notice that in the last two lines,
we can factor out a $u^2$ so that they can be recast as
\ba
-\tfrac{z^{-d}}2\,u^{2}&\Big[&
4\left[4(d+2s-6)+u^2\,\partial_u^2\right] {\mathcal O}+
(d+2s-5)(d+2s-7)\,\mathsf P_{\rm o}+ \nn
&&+\,(d+2s-5)\,
\mathsf P'_{\rm o}\,
\Big\{\,2(d+2s-3)-(d+2s-8)(3d+8s-10)\nn
&& \hspace{100pt}+\,4\,u^2\,\partial_u^2
-4(d+2s-7)\,\partial_u^2\,\mathsf T\, u^2\,\Big\}\,\Big]\,C^{\sst (s-4)}\,,
\label{u^{2}}
\ea
where we have used the definition of $\mathsf T$ as
well as the identities presented in Appendix \ref{sec: IDs}.
It is important to note that
\eqref{u^{2}} can be cancelled with a suitable choice of the operator $\mathcal O$\,.
In the latter case, the anomaly, that is the gauge variation of the action under $\epsilon^{\sst (s-2)}$\,, is given by
\ba
    \mathscr A^{\sst (s-2)} \e
    -\,\tfrac{z^{-d}}2\,\mathsf P_{\rm o}\,\partial_{u}^{2}\,A^{\sst (s)}
     -\,z^{-d+2}\,\Big\{\,\mathsf P_{\rm o}\left[\,
    \tfrac14(u\cdot\partial)^{2}-u\cdot\partial
    \left(u\cdot\partial\,\partial_u^2+3\,\partial_{u}\cdot\partial\right)\mathsf T\,u^{2}\,\right]\nn
    &&\hspace{50pt}
    +\,2\,(3\,u\cdot\partial+u^2\,\partial_u\cdot\partial)\,\mathsf T
    (4\,u\cdot\partial+u^2\,\partial_u\cdot\partial)\,
    {\mathcal O}\,\Big\}\,C^{\sst (s-4)}\,,
    \label{anom}
\ea
where the explicit form of $\mathcal O$ can be determined by
requiring \eqref{u^{2}} to be zero.
From the definitions of $A^{\sst (s)}$ and $C^{\sst (s-4)}$\,, one can see that the anomaly contains only terms proportional
to $z^{-d+2}$ or $z^{-d+3}$\,:
\be
    \mathscr A^{\sst (s-2)}=z^{-d+2}\,\mathcal A_{\phi}\,\phi^{\sst (s)}
    +z^{-d+3}\,\mathcal A_{\psi}\,\psi^{\sst (s-3)}\,.
\label{anom 24}
\ee
The absence of the terms proportional to $z^{-d}$ is due to the cancellation of \eqref{u^{2}},
and this will have important consequences later in analyzing the anomaly of the finite par of the on-shell action.

Finally, the total action is given by
\ba
 && \mathcal{I}_{\sst\mathcal M}+
 \mathcal{I}_{\sst \partial \mathcal M}+
  \mathcal{I}^{\sst\rm\,(c.t.)}_{\sst \partial \mathcal M} = \nn
&&=-\int^{\infty}_{z_{B}}dz\ \tfrac{z^{-d-1}}2\,
    \ppd{\varphi}{
    \left(1-\tfrac14\,U^{2}\,\partial_{U}^{2} \right) \mathcal{F}\,\varphi} \nn
&&\quad +\, z_{B}^{-d}\,
     \Big[\
    \tfrac12\,\ppd{ \phi^{\sst (s)}}{\mathsf{P}_{\rm e}'\,\chi^{\sst (s)}}_{z_{B}}
    +\tfrac{1}{4}\,
    \ppd{3\,\partial_u^{\,2}\,\phi^{\sst (s-1)}+\phi^{\sst (s-3)}}{
    \mathsf{P}_{\rm o}'\,\chi^{\sst (s-3)}}_{z_{B}}\nonumber\\
    &&\hspace{50pt} +\, \tfrac{1}{4}\,
    \ppd{\partial_u^{\,2}\,\phi^{\sst (s)}-\phi^{\sst (s-2)}}{
    \mathsf{P}_{\rm o}\,\zeta^{\sst (s-2)}}_{z_{B}} +
    \tfrac{1}{4}\,
    \ppd{\zeta^{\sst (s-2)}}{\mathsf P_{\rm o}''\,\zeta^{\sst
    (s,s-2)}}_{z_{B}}\nn
    &&\hspace{50pt} +\,
    \ppd{C^{\sst (s-4)}}{\mathcal O\ C^{\sst (s-4)}}_{z_{B}}
    +\ppd{\phi^{\sst (s)}}{G^{\sst (s)}}_{z_{B}}\!
    +\ppd{\beta^{\sst(s-3)}}{K^{\sst (s-3)}}_{z_{B}} \, \Big]\,.
\ea

\section{On-shell action}
\label{sec: OA}

The AdS/CFT corespondance, in the semi-classical regime, is the
identification between the finite parts of the effective action $W_{\Lambda}[h]$  of
the CFT and the on-shell HS action $S_{z_{B}}[h]$\,. The effective
action, a functional of a spin $s$ field $h^{\sst (s)}$, is
invariant under the gauge symmetry
$\delta\,h^{\sst (s)}=u\cdot\partial\,\lambda^{\sst (s-1)}$ and may be anomalous under the
conformal transformation $\delta\,h^{\sst (s)}=-u^2\,\sigma^{\sst (s-2)}$\,.
In this section, we shall use our previous results in order to evaluate the
HS action when the field is on-shell.

First, we have to make precise the AdS boundary conditions and how
they are given in terms of $h^{\sst (s)}$. This not straightforward, since
as we saw, in the bulk of AdS we have the field $\phi^{\sst (s-r)}$,
with $r=0,1,2,3$ subject to the gauge transformations \eqref{decomp. bd necessaries}.
For $s=2$\,, the standard procedure is to fix the gauge to the radial one, that
is $\phi^{\sst (1)}=\phi^{\sst (0)}=0$ and to identify $h^{\sst (2)}$ as the
boundary value of $\phi^{\sst (2)}$ at the origin. This prescription
is not easily generalized to higher spins. In fact, the gauge
parameters can set to zero only $\phi^{\sst (s-1)}$ and
$\phi^{\sst (s-2)}$ but not the remaining one $\phi^{\sst (s-3)}$. So
we have to proceed differently.

Using the gauge transfomations {\it and} the equations of motion,
it is possible to reach a Coulomb-like (or synchronous-like)
gauge, that is
\be
    \partial_u^2\,\phi^{\sst (s)}=\partial_u\cdot\partial\,\phi^{\sst (s)}=0\,,\quad
    \phi^{\sst (s-1)}=\phi^{\sst (s-2)}=\phi^{\sst (s-3)}=0\,.
\ee
In this gauge, the equations of motion simplify considerably and reduce to
\be
    \left[(z\,\partial_{z})^{2}-d\,z\,\partial_{z}-m_{s}^{2}-s
     +z^{2}\,\Box\right]\phi^{\sst (s)}=0\,,
\ee
which can be solved as
\be
     \phi^{\sst (s)}(z,x)=
     z^{2-s}\,U_{\frac
     {d+2s-4}2}\big(z\sqrt{-\Box}\,\big)\,h^{\sst (s)}_{\sst\rm TT}(x)\,,
\ee
where $h^{\sst (s)}_{\sst\rm TT}$ is a transverse and traceless $d$-dimensional spin $s$ field :
\be
     \partial_u^2\ h^{\sst (s)}_{\sst\rm TT}
     =0\,,\qquad
     \partial_u\cdot\partial\ h^{\sst (s)}_{\sst\rm TT}=0\,,
\ee
and the Bessel-like function $U_{\nu}(z)$ is defined in \eqref{U def} of
Appendix \ref{sec: IDs}.
So we obtain the solution as a function of the boundary field
$h^{\sst (s)}_{\sst\rm TT}$\,. The latter cannot be identified directly with $h^{\sst (s)}$
because of the transverse and traceless properties of $h^{\sst (s)}_{\sst\rm TT}$\,.
Since $h^{\sst (s)}$ has a gauge invariance, we can identify it with
$h^{\sst (s)}_{\sst\rm TT}$ up to a gauge and conformal transformation as
\be
    h^{\sst (s)}_{\sst\rm TT}=
    h^{\sst (s)}-u\cdot\partial\ \bar\rho^{\sst (s-1)}[h^{\sst (s)}]
    +u^2\,\rho^{\sst (s-2)}[h^{\sst (s)}]\,,\label{tt}
\ee where
the traceless $\bar\rho^{\sst (s-1)}[h^{\sst (s)}]$ and
$\rho^{\sst (s-2)}[h^{\sst (s)}]$ are given in terms
of the trace and the divergence of $h^{\sst (s)}$ as is shown in the
Appendix \ref{sec: TT app}.

The general solution can now be written by performing a general gauge transformation as
 \ba
    && \phi^{\sst (s)}=z^{2-s}\,U_{\frac
     {d+2s-4}2}\big(z\sqrt{-\Box}\,\big)\,h^{\sst (s)}_{\sst\rm TT}
    +z\,(u\cdot\partial )\,\epsilon^{\sst (s-1)}
    -u^{2}\,\epsilon^{\sst (s-2)}\,,\nn
    && \phi^{\sst (s-1)}
    =(z\,\partial_{z}+s-1+u^{2}\,\partial_{u}^{2})\,
    \epsilon^{\sst (s-1)}
    +z\,(u\cdot\partial )\,\epsilon^{\sst (s-2)}\,,\nn
    && \phi^{\sst (s-2)}
    =-z\,(u\cdot\partial )\,\partial_{u}^{2}\,\epsilon^{\sst (s-1)}
    +\left[2\left(z\,\partial_{z}+s-2\right)+u^{2}\,\partial_{u}^{2}\right]
    \epsilon^{\sst (s-2)}\,,\nn
    && \phi^{\sst (s-3)}
    =-\left[3\left(z\,\partial_{z}+s-3\right)+u^{2}\,\partial^{2}_{u}\right]
    \partial_{u}^{2}\,\epsilon^{\sst (s-1)}
    -z\,(u\cdot\partial )\,\partial_{u}^{2}\,\epsilon^{\sst (s-2)}\,.
    \label{sol tt}
\ea
Now, if we choose the gauge parameters as
\ba
    && \epsilon^{\sst (s-1)}=z^{1-s}\,U_{\frac
     {d+2s-4}2}\big(z\sqrt{-\Box}\,\big)\,\bar\rho^{\sst (s-1)}[h^{\sst (s)}]\,,\nn
    && \epsilon^{\sst (s-2)}=z^{2-s}\,U_{\frac
     {d+2s-4}2}\big(z\sqrt{-\Box}\,\big)\,\rho^{\sst (s-2)}[h^{\sst (s)}]\,,
 \label{ep rho}
\ea
then the solution can be expressed in terms of $h^{\sst (s)}$ as
\ba
    && \phi^{\sst (s)}=z^{2-s}\,U_{\frac
     {d+2s-4}2}\big(z\sqrt{-\Box}\,\big)\,h^{\sst (s)}\,,\nn
    && \phi^{\sst (s-1)}
    =z^{2-s}\,\partial_{z}\,
    U_{\frac
     {d+2s-4}2}\big(z\sqrt{-\Box}\,\big)\,\bar\rho^{\sst (s-1)}[h^{\sst (s)}]
    +z^{3-s}\,U_{\frac
     {d+2s-4}2}\big(z\sqrt{-\Box}\,\big)\,u\cdot\partial\,\rho^{\sst (s-2)}[h^{\sst (s)}]\,,\nn
    && \phi^{\sst (s-2)}
    =z^{2-s}\left(2\,z\,\partial_{z}+u^{2}\,\partial_{u}^{2}\right)
    U_{\frac{d+2s-4}2}\big(z\sqrt{-\Box}\,\big)\,\rho^{\sst (s-2)}[h^{\sst (s)}]\,,\nn
    && \phi^{\sst (s-3)}
    = -z^{3-s}\,U_{\frac{d+2s-4}2}\big(z\sqrt{-\Box}\,\big)\,
    u\cdot\partial\,\partial_{u}^{2}\,\rho^{\sst (s-2)}[h^{\sst (s)}]\,,
    \label{solu}
\ea
where we used the tracelessness of $\bar\rho^{\sst (s-1)}[h^{\sst (s)}]$\,.
Now, our proposition for $S[h^{\sst (s)}]$ is to evaluate
$\mathcal I[\varphi]=\mathcal I[\phi^{\sst (s)},\phi^{\sst (s-1)},\phi^{\sst (s-2)},\phi^{\sst (s-3)}]$
on the above solution which generates a functional of $h^{\sst (s)}$\,.

\subsection{Symmetries and anomalies of the on-shell action}

In this and the  following subsections we examine the properties of
the on-shell action:
\be
    S_{z_{B}}[h^{\sst (s)}] :=
    \big(\,\mathcal I_{\sst \mathcal M}+\mathcal I_{\sst \partial \mathcal M}
    + \mathcal I_{\sst \partial\mathcal M}^{\sst\,\rm{(c.t.)}}\,\big)
    \big[\,\phi^{\sst (r)}[h^{\sst (s)}]\,\big]\,.
\ee
Let us first note  that it has an
expansion in $z_B$\,:
\be
S_{z_{\sst B}}=
\sum_{n=3-s}^{[d/2]}
z_{\sst B}^{2n-d}\,S_{n} + S_{\rm fin} +
\log(\mu\,z_{\sst B})\,S_{\rm log}+o(z_B)\,,
\label{S exp}
\ee
where the logarithmic terms are given with a constant $\mu$ of mass dimension
and they are absent for odd dimensions,
$o(z_{\sst B})$ vanishes for small $z_{\sst B}$ faster than $z_{\sst B}$\,.
Of particular importance are the finite (or renormalized) part $S_{\rm fin}$ and logarithmic part $S_{\rm log}$\,.
The CFT effective action $W_\Lambda$ has a similar expansion in terms of
the cut-off $\Lambda$, the renormalized part $W_{\rm fin}$ should, by the
holographic correspondence, be equal (up to possible local counterterms)
to $S_{\rm fin}$\,.
The AdS isometry $(z',x')=(\Omega\,z, \Omega\,x)$, under
which $h^{\sst (s)}$ transforms as
\be
    h^{\sst (s)}(x)\quad \to \quad
    h'^{\sst(s)}(x)=\Omega^{2-s}\,h^{\sst (s)}(\Omega^{-1}\,x)\,,
\ee
implies that $S_{\Omega\,z_{B}}[h'^{\sst (s)}]=S_{z_B}[h^{\sst (s)}]$\,.
Combined with \eqref{S exp}, this in turn gives the dilatation
transformations of $S_{n}$ as
\be
    S_{n}[h'^{\sst(s)}]=\Omega^{2n-d}\,S_{n}[h^{\sst (s)}]\,.
\ee
For odd $d$\,, the relation holds also for $S_{\rm fin}$\,, whereas for even $d$ where
logarithmic terms are also present we get
\be
    S_{\rm log}[h'^{\sst(s)}]=S_{\rm log}[h^{\sst (s)}]\,,\qquad
    S_{\rm fin}[h'^{\sst (s)}]+\log\Omega\,S_{\rm log}[h'^{\sst (s)}]=S_{\rm fin}[h^{\sst (s)}]\,.
\ee

Next, let us examine the transformation under the gauge
symmetries. Under a gauge transformation $\delta\,h^{\sst
(s)}=u\cdot\partial\,\bar\lambda^{\sst (s-1)}
-u^{2}\,\sigma^{\sst (s-2)}$\,, with arbitrary $\sigma^{\sst
(s-2)}$ and traceless $\bar\lambda^{\sst (s-1)}$\,, we have \be
    \delta\,\bar\rho^{\sst (s-1)}[h^{\sst (s)}]=\bar\lambda^{\sst (s-1)}\,,\qquad
    \delta\,\rho^{\sst (s-2)}[h^{\sst (s)}]=\sigma^{\sst (s-2)}\,.
\ee Hence, from \eqref{ep rho} one can find that the on-shell
action is invariant under $\bar\lambda^{\sst (s-1)}$ as it should,
while under a conformal transformation by $\sigma^{\sst (s-2)}$,
we have \be
     \delta_{\sigma}\,S_{z_{B}}
     =\ppd{\epsilon^{\sst (s-2)}}{\mathscr A^{\sst
     (s-2)}}_{z_{B}}\,,\label{anoh}
\ee
where $\epsilon^{\sst (s-2)}=z^{2-s}\,
U_{\frac{d+2s-4}2}\big(z\sqrt{-\Box}\,\big)\,\sigma^{\sst (s-2)}$ and
$\mathscr A^{\sst (s-2)}$ \eqref{anom}
is now expressed in terms of $h^{\sst (s)}$\,.
Its explicit expression is rather cumbersome, let us just examine
the $z_{\sst B}$ dependence of $\delta_{\sigma}\,S_{z_{B}}$. In order to do that, we need the explicit
dependence of $\mathscr A^{\sst(s-2)}$ on $z_{\sst B}$ which we write,
from \eqref{anom 24} and \eqref{solu}, as
\be
     \mathscr A^{\sst (s-2)}=z_{\sst B}^{-d-s+4}\,
     U_{\frac{d+2s-4}2}\big(z_{\sst B}\sqrt{-\Box}\,\big)\,
     \big(\mathscr A^{\sst (s-2)}_{\phi}+z_{\sst B}^{2}\,
     \mathscr A^{\sst (s-2)}_{\psi}\big)\,,
\ee
where $\mathscr A^{\sst (s-2)}_{\phi}$ and $\mathscr A^{\sst (s-2)}_{\psi}$
are given from \eqref{anom 24} by
\be
     \mathscr A^{\sst (s-2)}_{\phi}=\mathcal A_{\phi}\,h^{\sst (s)}\,,
     \qquad
     \mathscr A^{\sst (s-2)}_{\psi}=
     \tfrac12\,\mathcal A_{\psi}\,
     (3\,\partial_{u}\!\cdot\partial+u\cdot\partial\,\partial_{u}^{2})\,
     \rho^{\sst(s-2)}[h^{\sst (s)}]\,.
\label{anom z}
\ee
Collecting all terms depending on $z_B$ in (\ref{anoh}) we get
\be
    \delta_{\sigma}\,S_{z_{B}}=
    z_{\sst B}^{-d-2s+6}\,
    \ppd{\sigma^{\sst (s-2)}}
      {\big[\,U_{\frac{d+2s-4}2}\big(z_{\sst B}\sqrt{-\Box}\,\big)\,\big]^2\,
      \big(\mathscr A^{\sst (s-2)}_{\phi}+z_{\sst B}^{2}\,
      \mathscr A^{\sst (s-2)}_{\psi}\big)}\,.
 \ee
Using the expansion of $U_\nu(z)$ given in Appendix \ref{sec: IDs},
 we see that for odd  $d$ (that is for half integer $\nu$) there is
 no contribution proportional to $(z_{\sst B})^0$\,, similarly for even $d$ (or integer
 $\nu$) there is no  contribution proportional to $\log(\mu\,z_{\sst B})$\,.
 We conclude that
 \be
\delta_{\sigma}\,S_{\rm fin}=0\quad [\,d:{\rm odd}\,]\,,
\qquad
\delta_{\sigma}\,S_{\log}=0\quad
[\,d:{\rm even}\,]\,.
\ee
Notice that this property is directly related
to the counterm which removed the $z^{-d}$-independent part of the
anomaly $\mathscr A^{\sst (s-2)}$ \eqref{anom}.

The explicit expression for the anomaly (\ref{anom z}) in terms of
$h^{\sst (s)}$ simplifies for low spins, for instance the   $s=3$
anomaly is easily  obtained after we first determine $\rho^{\sst (1)}[h^{\sst (3)}]$ as
\be
\rho^{\sst (1)}[h^{\sst (3)}]=-\tfrac{1}{4(d+1)\,\Box}\left(\,
1-\tfrac{d-2}{3d}\tfrac{u\cdot\partial\,\partial_u\!\cdot\,\partial}
{\Box}\,\right)\partial_u^2\,\mathcal F_{M^{d}}\,h^{\sst (3)}\,,
\ee
the anomaly is then given by
\be
    \mathscr A^{\sst (1)}_{\phi}=   -\tfrac{1}{2}\,
\partial_u^2\,\mathcal F_{M^{d}}\,h^{\sst (3)}\,,
\qquad
  \mathscr A^{\sst (1)}_{\psi}=
   -\tfrac1{4d(d+1)}\,
     u\cdot\partial\,\partial_u\!\cdot\partial\,
\partial_u^{\,2}\,\mathcal F_{M^{d}}\,h^{\sst (3)}\,.
\ee

\subsection{Spin 2 case}
In this subsection, we give the explicit expression for the
on-shell action as a functional of the conformal field $h^{\sst (2)}$\,. We
then compare it with the on-shell action computed in the radial
gauge. The two on-shell actions, as we will see,
coincide.

As we showed in the previous Section, the addition of a local
couterterm allows the recovery of the full bulk gauge symmetry if
$d> 2$\,. When expresssed in terms of the transverse and traceless
field $h_{\rm\sst TT}^{\sst (2)}$,
the action reduces to
\be
S_{z_{B}}=
-\tfrac{z_B^{-d}}{2}\,
\ppd{\phi^{\sst (2)}[h^{\sst (2)}_{\rm\sst TT}]}
{z\,\partial_{z}\,\phi^{\sst (2)}[h^{\sst (2)}_{\rm\sst TT}]}_{z_{B}}
+\tfrac{z_B^{-d+2}}{2(d-2)}\,
\ppd{\phi^{\sst (2)}[h^{\sst (2)}_{\rm\sst TT}]}
{\Box\,\phi^{\sst (2)}[h^{\sst (2)}_{\rm\sst TT}]}_{z_{B}}\,,
\ee
where the traces and divergences of $\phi^{\sst (2)}[h_{\rm\sst TT}^{\sst (2)}]$ vanish
so do not contribute, and the second term comes from the counterterm \eqref{c.t. 2}.
Using the explicit form of solutions \eqref{sol tt} and the identity \eqref{U id}, one gets
\be
S_{z_{B}}=-
\tfrac{z_B^{-d+4}}{2(d-2)}\,
\ppd{h^{\sst (2)}_{\rm\sst TT}}
{V_{\frac{d-2}2}\big(z_{\sst B}\sqrt{-\Box}\,\big)\,
\Box^{2}\,
h^{\sst (2)}_{\rm\sst TT}}\,,
\ee
with
\be
    V_{\nu}(z):=\tfrac1{z^{2}}\,
    U_{\nu+1}(z)\left[U_{\nu}(z)-U_{\nu+1}(z)\right]=
    -\tfrac{1}{4\,\nu(\nu-1)}\,
    U_{\nu+1}(z)\,U_{\nu-1}(z)\,.
\ee
The expression of $h^{\sst (2)}_{\rm\sst TT}$ in terms of $h^{\sst (2)}$
can be deduced as the general case presented in
the Appendix \ref{sec: TT app}, or alternatively by applying first the Fronsdal
operator to the definition \eqref{tt} to get
\be
    \mathcal F_{M^{d}}\,h^{\sst (2)}_{\rm\sst TT}
    =\mathcal F_{M^{d}}\,h^{\sst (2)}
    +\left[u^2\,\Box+(d-2)(u\cdot\partial)^2\right]
    \rho^{\sst (0)}[h^{\sst (2)}]\,,
\ee
and then taking the trace of the equation to determine $\rho^{\sst (0)}[h^{\sst (2)}]$ as
\be
\rho^{\sst (0)}[h^{\sst (2)}]=-\tfrac{1}{4(d-1)\,\Box}\,\partial_u^2\,\mathcal F_{M^{d}}\,
h^{\sst (2)}\,.
\ee
Similarly, $\rho^{\sst (1)}[h^{\sst (2)}]$ can be obtained by first taking the trace of
\eqref{tt} which determines $\partial_u\cdot\partial\,\rho^{\sst (1)}[h^{\sst (2)}]$
and then the divergence of \eqref{tt} to get
\be
\rho^{\sst (1)}[h^{\sst (2)}]=\tfrac1{\Box^2}
\left[\Box\,
\partial_{u}\!\cdot\partial -\tfrac1{2(d-1)}\,\Box\,
u\cdot\partial\,\partial_u^2 -\tfrac{d-2}{2(d-1)}\,u\cdot\partial\,
(\partial_u\!\cdot\partial
)^2\right] h^{\sst (2)}\,.
\ee
Finally , $h^{\sst (2)}_{\rm\sst TT}$ is expressed in terms of
$h^{\sst (2)}$ in the manifestly gauge invariant way as
\be
h^{\sst (2)}_{\rm\sst TT}
=\mathcal P^{\sst (2)}_{\rm\sst TT}\,h^{\sst (2)}\,,
\qquad
\mathcal P^{\sst (2)}_{\rm\sst TT}
=\left[1-\tfrac{1}{4(d-1)}u^2\,\partial_u^2-
\tfrac{d-2}{4(d-1)}\tfrac{(u\cdot\partial)^2}{\Box}\partial_u^2\right]
\tfrac{\mathcal F_{M^{d}}}{\Box}\,,
\ee
where $\mathcal P^{\sst (2)}_{\rm\sst TT}$ is
the traceless and transverse projector of spin 2 field.
Finally, the on-shell action as a functional of $h^{\sst (2)}$ can be deduced as
 \be
 S_{z_{B}}= -\tfrac{z_B^{-d+4}}{2(d-2)}\,
 \ppd{h^{\sst (2)}}{
 V_{\frac{d-2}2}\big(z_{\sst B}\sqrt{-\Box}\,\big)\,\mathcal C^{\sst (2)}\, h^{\sst (2)}}\,,
  \label{OA 2}
 \ee
where $\mathcal C^{\sst (2)}$ is given by a local differential operator:
\be
    \mathcal C^{\sst (2)}:=\Box^{2}\  {\mathcal P^{\sst (2)}_{\sst\rm TT}}\,.
\ee
Notice that, for $d\ge 4$\,, this result can be re-expressed in terms of the
Weyl tensor $\mathscr C^{\sst (2)}$
by making use of the identity:
\be
    \ppd{\mathscr C^{\sst (2)}}{\mathscr C^{\sst (2)}}   =  6\,\tfrac{d-3}{d-2}\,\ppd{h^{\sst (2)}}{\mathcal C^{\sst (2)}\,h^{\sst (2)}}\,.
\ee

We now compare with the  radial gauge treatment which is valid for
spin 2. The solution can be readily determined  as
\be
    \phi^{\sst (2)} =
    U_{\frac d2}\big(z\sqrt{-\mathcal H}\,\big)\,h^{\sst
    (2)}\,,\quad
    \phi^{\sst (1)} =\phi^{\sst (0)}=0\,,
    \label{spin 2 rad sol}
\ee where \be
    \mathcal H:= \left(1-\tfrac1{4\,(d-1)}\,u^{2}\,\partial_{u}^{2}\right)
    \mathcal{F}_{M^{d}}\,.
\ee
The operator $\mathcal H$ satisfies \be
    (\partial_{u}\!\cdot\partial-u\cdot\partial \,\partial_{u}^{2})\,\mathcal H=0\,,\nn
    \qquad \partial_{u}\!\cdot\partial\ \mathcal H^{2}=0\,,
    \qquad
    \partial_{u}^{2}\ \mathcal H^{2}=0\,,
    \label{H iden}
\ee and, as a result: \be
    \mathcal{H}^{n+2}=\Box^{n}\,\mathcal{H}^{2}\,,\qquad
    [n\in \mathbb{N}]\,.
    \label{H iden2}
\ee
 The on-shell action with the radial gauge solution
\eqref{spin 2 rad sol} can now be determined as
\ba
   S_{z_{B}} \e
     -\tfrac{z_{B}^{-d}}2\,
    \ppd{ \phi^{\sst (2)}[h^{\sst (2)}]}{
    \left(1-\tfrac 12\,u^{2}\,\partial_{u}^{2}\right)
    z\,\partial_{z}\, \phi^{\sst (2)}[h^{\sst (2)}]}_{z_{B}} \nn
    &&
    +\,\tfrac{z_B^{-d+2}}{2(d-2)}\,
    \ppd{\phi^{\sst (2)}[h^{\sst (2)}]}{\left(1-\tfrac{1}{4}\,u^{2}\,\partial_u^2\right)
    {\cal F}_{M^{d}}\,\phi^{\sst (2)}[h^{\sst (2)}]}_{z_{B}}\,.
\ea
By making use of the $z$-derivative of the solution:
\be
    z\,\partial_{z}\,\phi^{\sst (2)}
    =\tfrac1{d-2}\,z^{2}\,\mathcal H\,U_{\frac {d-2}2}\big(z\sqrt{-\mathcal{H}}\,\big)\,
    h^{\sst (2)}\,,
    \label{z p2}
\ee
and the identity \eqref{H iden2} and
\be
    \ppd{\mathcal F_{M^{d}}\,h^{\sst (2)}}{
    \left(1-\tfrac{d}{8(d-1)}\,u^{2}\,\partial_{u}^{2}\right)
    \mathcal F_{M^{d}}\,h^{\sst (2)}}
    = \ppd{h^{\sst (2)}}{\mathcal C^{\sst (2)}\, h^{\sst (2)}}\,,
\ee
one can show that
the on-shell action agrees with \eqref{OA 2}.

Notice that this on-shell action \eqref{OA 2} is invariant under gauge and Weyl symmetry,
\m{\delta\,h^{\sst (s)}=u\cdot\partial\,\lambda^{\sst (s-1)}-u^{2}\,\sigma^{\sst (s-2)}}\,, and
notice also that  all the divergent
terms in the small $z_{\sst B}$ limit are local. Of particular interest is
the finite term for odd $d$ and the logarithmical divergent term for even $d$\,.
Using the expansion of $U_\nu(z)$ given in Appendix \ref{sec: IDs},
the finite part of the on-shell action for odd $d$ is given by
\be
    S_{\rm fin}=
    \frac{\pi\,(-1)^{\frac{d+1}2}}{2^{4}\,\Gamma^2(\tfrac d2)
    }\,
    \ppd{h^{\sst (2)}}
    {\left(-\tfrac14\,\Box\right)^{\frac{d-4}2}
 \mathcal C^{\sst (2)}\, h^{\sst (2)}}\,,
    \label{s2 onac odd}
\ee and the logarithmically divergent part for even $d$ by
\be
    S_{\rm log}=
    \frac{(-1)^{\frac{d+2}2}}{2^{3}\,\Gamma^{2}(\tfrac d2)}\,
    \ppd{ h^{\sst (2)}}
    {\left(-\tfrac14\,\Box\right)^{\frac{d-4}2}
  \mathcal C^{\sst (2)}\,
    h^{\sst (2)}}\,.
    \label{s2 onac even}
\ee

\subsection{General case}

The on-shell action reduces, as before,  to the boundary term
$\mathcal I_{\sst \partial \cal M}+
    \mathcal I^{\sst {\,\rm (c.t.)}}_{\sst \partial \cal M}$
(\ref{boundary})
where the fields
should be replaced with the solution (\ref{solu}). The $z_{\sst B}$
expansion of the action is obtained from the expansion of
$U_\nu(z)$. The latter is of the form:
{
\be\label{Uexp}
    U_\nu(z)=
    P_\nu(z^2)+z^{2\,\nu} \left\{
    \begin{array}{cc}
   (-1)^{-\frac12}\, \frac\pi2\ Q_\nu(z^2)\qquad &[\nu\in \mathbb N+\tfrac12] \\
    \log z \,Q_\nu(z^2)+R_{\nu}(z^{2}) \qquad &[\nu\in\mathbb N]
    \end{array}\right.
\ee
where $P_{\nu}$, $Q_\nu$ and $R_\nu$ are power series.}

The HS-Weyl-invariant part of the on-shell action,
that is $S_{\rm fin}$ for odd $d$ and $S_{\log}$ for even $d$\,, can be deduced
easily from the one evaluated with the $h^{\sst (s)}_{\sst\rm TT}$ solution \eqref{sol tt}\,:
 \be
 S_{z_{B}}[h^{\sst (s)}_{\sst\rm TT}]=
 \frac{z_{\sst B}^{-d-2s+6}}{2(d+2s-6)}\,
 \ppd{h^{\sst (s)}_{\sst\rm TT}}
{U_{\frac{d+2s-4}2}\big(z_{\sst B}\sqrt{-\Box}\,\big)\,
U_{\frac{d+2s-6}2}\big(z_{\sst B}\sqrt{-\Box}\,\big)\,
\Box\,h^{\sst (s)}_{\sst\rm TT}}\,.
\label{OA s}
\ee
{One can see from \eqref{Uexp} that 
the finite part or logarithmically divergent part of \eqref{OA s} is given by
the $Q_{\frac{d+2s-4}2}(0)\,P_{\frac{d+2s-6}2}(0)$  term.}
The finite part for odd $d$ is
\be
S_{\rm fin}[h^{\sst (s)}_{\sst\rm TT}]=S_{\rm fin}[h^{\sst (s)}]
=\frac{\pi\, (-1)^{\frac{d+2s-3}2}}{2^{2s}\,\Gamma^2(\frac{d+2s-4}2)}\,
\ppd{h^{\sst (s)}}
{ \left(-\tfrac14\,\Box\right)^{\frac{d-4}2}\,\mathcal C^{\sst (s)}\,
    h^{\sst (s)}}\,,
\label{OA fin}
\ee
and a similar expression for the logarithmically divergent term for even $d$\,:
\be
S_{\log}[h^{\sst (s)}_{\sst\rm TT}]=S_{\log}[h^{\sst (s)}]
=\frac{(-1)^{\frac{d+2s-2}2}}{2^{2s-1}\,\Gamma^2(\frac{d+2s-4}2)}\,
\ppd{h^{\sst (s)}}
{ \left(-\tfrac14\,\Box\right)^{\frac{d-4}2}\,
\mathcal C^{\sst (s)}\,h^{\sst (s)}}\,,
\label{OA log}
\ee
where $\mathcal C^{\sst (s)}$ is a local differential operator given
with the traceless and transverse projector
$\mathcal P^{\sst (s)}_{\rm\sst TT}$ \eqref{P tt} by
\be
    \mathcal C^{\sst (s)}:={(-\Box)^{s}}\,\mathcal P_{\rm\sst TT}^{\sst (s)}\,.
\ee
To derive the above results, we have used the fact that $\delta_{\sigma}\,S_{\rm fin}=0$ for odd $d$
and $\delta_{\sigma}\,S_{\log}=0$ for even $d$\,.
As in the spin 2 case, these results can be re-expressed for $d\ge 4$
in terms of the spin $s$ Weyl tensor $\mathscr C^{\sst (s)}$ since
\be
      \ppd{\mathscr C^{\sst (s)}}{\mathscr C^{\sst (s)}}
      \propto
    \ppd{h^{\sst (s)}}{\mathcal C^{\sst (s)}\,h^{\sst (s)}}\,.
    \label{cont Weyl}
\ee
See Appendix \ref{sec: TT app} for more details.

For low spins the explicit expression is easily obtained.
We give here the complete expression for spin $3$
in a manifestly gauge invariant way. The projector $\mathcal P_{\sst\rm TT}^{\sst (3)}$
is given by
\be \mathcal P^{\sst (3)}_{\sst\rm TT}=\Big[1-
\tfrac{d}{4(d+1)}\tfrac{(u\cdot\partial)^2}{\Box}\,\partial_u^2 +
\tfrac{d-2}{12(d+1)}\tfrac{(u\cdot\partial)^3\partial_u\cdot\partial}{\Box^2}
\partial_u^2
-\tfrac{1}{4(d+1)}u^2\,\partial_u^2+
\tfrac{1}{4(d+1)}u^2\,
\tfrac{u\cdot\partial\partial_u\cdot\partial}{\Box}\,\partial_u^2\Big]\,
\tfrac{\mathcal F_{M^{d}}}{\Box}\,, \ee
and the on-shell action can be calculated using the
solution (\ref{sol tt}) as
\ba\label{OA s3}
&&
S_{z_{B}}=
\tfrac{z_B^{-d}}{2d}\,\ppd{h^{\sst (3)}}{
U_{\frac{d+2}2}\big(z_{\sst B}\sqrt{-\Box}\,\big)\,
U_{\frac{d}2}\big(z_{\sst B}\sqrt{-\Box}\,\big)\,\tfrac1{\Box^{2}}\,
\mathcal C^{\sst (3)}\,h^{\sst (3)}} \nn
&&+\,\tfrac{z_B^{-d+2}}{8(d+1)}\,
\ppd{\partial_u^2\,\mathcal F_{M^{d}}\,h^{\sst (3)}}
{\Big[\,U_{\frac{d+2}2}\big(z_{\sst B}\sqrt{-\Box}\,\big)\,\Big]^2
\Big(1-\tfrac{d-2}{3d}\,\tfrac{u\cdot\partial\,\partial_u\cdot\partial}{\Box}\Big)
\Big(1+\tfrac{z_B^2\,u\cdot\partial\,\partial_u\cdot\partial}{2d\,(d+1)}\Big)\,
\partial_u^2\,\mathcal F_{M^{d}}\, h^{\sst(3)}}\,. \nn
\ea
The first line is invariant under the
conformal transformation and the second line reproduces the
anomaly.

{Notice that the on-shell action $S_{z_{B}}$ \eqref{OA s3} is given by a linear combination
of tensor structures whose coefficients are functions of the form
$ U_{\nu}(z)\,U_{\nu'}(z)$ with $z=z_{B}\sqrt{-\Box}$\,.
In the expansion of the latter, 
only the terms compensating the overall power of $z_{B}$
contribute to the finite or logarithmically divergent part. 
This pattern is general for arbitrary spins, and 
one can see from \eqref{Uexp} that
$S_{\rm fin}$ (odd $d$) or  $S_{\rm log}$ (even $d$) is determined by
the $(-1)^{-\frac12}\,\frac\pi2\ z^{2\nu}\,Q_{\nu}\,P_{\nu'}$ or 
$z^{2\nu}\,\log z\,Q_{\nu}\,P_{\nu'}$ term respectively.
Therefore, they share the same tensor structure differing only by a factor of 
$(-1)^{-\frac12}\,\frac\pi2$\,, and they are both invariant as we have shown in this section. 
In the case of $S_{\rm fin}$ for even $d$\,, the situation is rather different
and the relevant term in the expansion of $U_{\nu}\,U_{\nu'}$ is not only
$z^{2\nu}\,\log z\,Q_{\nu}\,P_{\nu'}$
but also $P_{\nu}\,P_{\nu'}$ and $z^{2\nu}\,R_{\nu}\,P_{\nu'}$\,.\footnote{
The remaining terms $z^{2(\nu+\nu')}\,\log z\,Q_{\nu}\,R_{\nu'}$ and $z^{2(\nu+\nu')}(\log z)^{2}\,Q_{\nu}\,Q_{\nu'}$ contribute to the on-shell action
with terms which vanish in the limit $z_{B}\to 0$\,.}
The latter two terms being associated with different tensor structures
are responsible for the anomaly. However, they give local contributions to the on-shell action
since they do not involve any term with $\log\sqrt{-\Box}$ or odd powers of $\sqrt{-\Box}$\,.
Consequently, the non-local part of $S_{\rm fin}$ (even $d$) has a similar form as \eqref{OA log}
with an insertion of $\log\sqrt{-\Box}$\,.}

\section{Conclusion}
\label{sec: concl}

In this paper we have determined the on-shell action for a free
higher-spin gauge field in AdS. In order to accomplish this we
first determined the boundary terms which are necessary to
reproduce the equation of motion in the presence of a finite
distance boundary. We constrained the  boundary terms with the
requirement of invariance under the symmetry transformations of
the bulk equations of motion and showed that for spins larger than
two there is an anomaly: no local boundary term can reproduce all
the gauge symmetry of the bulk. Next, we have written the solution
of the bulk equations of motion in terms of a boundary higher-spin
conformal field $h^{\sst (s)}$, that is a $d$-dimensional field
subject to gauge and HS Weyl transformations. Using these two
building blocks, the boundary action and the boundary field, we
determined the on-shell action $S_{z_B}[h^{\sst (s)}]$.

The key motivation of our investigation is the HS/CFT
correspondence. Let us examine how the on-shell action is
described from the boundary CFT point of view.
 The boundary CFT is the theory of
$N$ free massless complex scalars,
$\bm\phi=(\phi^{1},\cdots,\phi^{N})$, in $d$ dimensions. The
operators dual to the AdS HS gauge fields are the currents of
higher ranks: \be
    \bar J^{{\sst (s)}\,\mu_{1}\cdots\mu_{s}}
    =\sum_{r=0}^{s}\,c_{r,s}\,\partial^{\mu_{1}}\cdots
    \partial^{\mu_{r}}\bm \phi^{*} \bm\cdot
    \partial^{\mu_{r+1}}\cdots \partial^{\mu_{s}}\bm \phi\,,
\ee where the coefficients $c_{r,s}$'s are chosen such that these
currents are conserved and traceless when the scalar is on shell:
\be
    \partial_{\mu_{1}}\,\bar J^{{\sst (s)}\,\mu_{1}\cdots\mu_{s}} \approx 0\,,
    \qquad
    \eta_{\mu_{1}\mu_{2}}\,\bar J^{{\sst (s)}\,\mu_{1}\cdots\mu_{s}} \approx 0\,.
    \label{cons. tr.}
\ee The connected correlation functions of these currents can be
generated by the effective action $W_{\Lambda}[\tilde h^{\sst
(0)},\tilde h^{\sst (1)},\cdots]$ given by \be
    e^{-{W}_{\Lambda}}:=
    \int_{\Lambda} \mathcal D\bm\phi\
    \exp\left(-\int d^{d}x\,\partial_{\mu}\bm\phi^{*}\bm\cdot\partial^{\mu}
    \bm\phi
    +\sum_{s=0}^{\infty} \ppd{\tilde h^{\sst (s)}}{\bar J^{\sst (s)}} \right),
    \label{eff act}
\ee where the (boundary) HS fields $\tilde h^{\sst (s)}$ is the
source for the currents $\bar J^{\sst (s)}$, and $\Lambda$ is the
UV regularization parameter of mass dimension. The effective
action $W_{\Lambda}$ \eqref{eff act} is a functional of the
conformal HS fields $\tilde h^{\sst (s)}$\,. According to the
HS/CFT ccorrespondence the finite part of $W_{\Lambda}$ is given,
in the semi-classical regime, by the finite part of the bulk
on-shell action. The traceless-ness and transversality of the
currents \eqref{cons. tr.} induce respectively the
diffeomorphism-like and the Weyl-transformation-like gauge
transformations on $\tilde h^{\sst (s)}$\,: \be
    \delta\, \tilde h^{\sst (s)}= u\cdot\partial\,\lambda^{\sst (s-1)}-
    u^{2}\,\sigma^{\sst (s-2)}+
    \mathcal O(\tilde h, \lambda) + \mathcal O(\tilde h, \sigma)\,,
    \label{bnd gt}
\ee where  $\mathcal O(\tilde h,\lambda)$ and $\mathcal O(\tilde
h,\sigma)$ are linear in $\tilde h^{\sst (s')}, \lambda^{\sst
(r)}$ and $\sigma^{\sst (r')}$ for any $s', r,r'$\,, and their
explicit expressions are given in \cite{Bekaert:2010ky}\,.

When $\tilde h^{\sst (s)}$ is restricted to only spin 2 field, the
transformations \eqref{bnd gt} reduce to the diffeomorphism and
the Weyl transformation.
 Here, it is important to note that the gauge field $\tilde h$ and the gauge parameters
$\lambda$ and $\sigma$ are not subject to any trace constraints,
and also that the deformations of the linear gauge transformations
are exact in the first order of $\tilde h$: there is no $\mathcal
O(\tilde h^{2})$ term. In \cite{Bekaert:2010ky}\,, it has been
shown that $W_{\Lambda}$ is invariant only under the
$\lambda$-transformations while the $\sigma$-transformations are
anomalous: $\delta_{\lambda}\, W_{\Lambda}=0$ and
$\delta_{\sigma}\,W_{\Lambda} \neq 0$\,. Further analysis was
performed  by expanding $W_{\Lambda}$ in $\Lambda$ as \be
    W_{\Lambda}=
    \sum_{n=-\infty}^{[d/2]} \Lambda^{-2n+d}\,W_{n}+
    W_{\rm fin}-\log\tfrac\Lambda\mu\,W_{\rm log}+
    o(\Lambda^{-1})\,,
    \label{W exp}
\ee
and it has been shown that
\be
    \delta_{\sigma}\,W_{\rm fin}=0
    \quad [\,d : {\rm odd}\,]\,, \qquad
    \delta_{\sigma}\,W_{\rm log}=0 \quad [\,d : {\rm even}\,]\,,
    \label{W even}
\ee which generalizes the case of the spin 2 Weyl anomaly. The
explicit expressions of $W_{\rm fin}$ and $W_{\log}$ were
determined perturbatively in the fields in
ref.\cite{Bekaert:2010ky}, the quadratic terms can be expressed in
terms of the on-shell action as
\ba
    &&-W_{\rm fin}[\tilde h]=\sum_{s=0}^{\infty}\,
    S_{\rm fin}[C_{s,d}\,\tilde h^{\sst (s)}]
    +\mathcal O(\tilde h^{3})
    \qquad [d : {\rm odd}]\,,\nn
    && -W_{\rm log}[\tilde h]=\sum_{s=0}^{\infty}\,
    S_{\log}[C_{s,d}\,\tilde h^{\sst (s)}]+\mathcal O(\tilde h^{3})
    \qquad  [d : {\rm even}]\,,
    \label{CFT W}
\ea
with
\be
    C_{s,d}^2=\frac{\Gamma^{2}(\frac{d+2s-4}2)}
    {\pi^{\frac{d-1}2}\,2^{d+s+1}\,\Gamma(\frac{d+2s-1}2)}\,.
\ee Hence, the quadratic parts of the CFT effective action
coincide with the on-shell actions $S_{\rm fin}$ and $S_{\log}$
(\ref{OA fin}\,,\,\ref{OA log}) up to  field redefinitions with
multiplicative constants.
{
The same is also true for the non-local part of $S_{\rm fin}$ for even $d$\,.}

In this paper, we have shown that the AdS counterpart of the HS
Weyl anomaly is due essentially to the fact that the boundary
term, which is necessary to obtain the classical HS EOM, cannot be
made invariant under the whole bulk gauge group. More precisely,
the on-shell action breaks the symmetries generated by the gauge
parameters $\epsilon^{\sst (s-2)}$ or
$\varepsilon_{d\mu_{1}\cdots\mu_{s-2}}$\,. At the quadratic level,
the finite part of the on-shell action is anomalous only for an
even dimensional boundary and in this case the anomaly is due to
local terms, {thus contributing with only contact terms to the two-point correlation function.
More precisely, the non-local part of the on-shell action is anomaly-free for any dimension 
and coincides with that of the effective action determined in \cite{Bekaert:2010ky}.
Consequently, the correlation function of two currents 
$\la \bar J_{\mu_{1}\cdots\mu_{s}}(x)\,\bar J_{\nu_{1}\cdots\nu_{s}}(x')\ra$ 
obtained from the on-shell action
by functional differentiation with respect to the conformal HS fields
coincides with the one obtained from the free boundary CFT
up to possible contact terms.
The way to extract the correlation functions from the effective action has been
shown in \cite{Bekaert:2010ky}.} 
The anomaly is expected to play a more important role
when considering higher orders since in the CFT side the
generalized Weyl anomalies arise  starting from cubic orders (see
\cite{Vasiliev:2011xf,Joung:2011ww} for recent constructions of
cubic interaction vertices of AdS HS fields). A key problem for
the future is to gain a better understanding of the HS
interactions in the metric form from the CFT effective action.

Another important problem is to see via the AdS/CFT correspondence
how the non-Abelian deformation of the gauge transformation is
realized in the bulk of AdS. In fact, it was shown in \cite{Bekaert:2009ud}
that the collection of all the HS fields can be grouped in a
Hermitian operator acting on the scalar field and that
the transformations (\ref{bnd gt}) arise from the leading terms of respectively
the Hermitian and anti-Hermitian transformations acting on the scalar fields.
The symmetry group of the CFT effective action is thus generated
by the algebra of Hermitian operators. An important issue is the
bulk counterpart of this symmetry group.
 A step in this direction was proposed by the correspondence which
we made explicit between the boundary conformal HS field and the bulk
HS field. We hope to come back to this issue in the future.

\acknowledgments{We are grateful to X. Bekaert and A. Sagnotti for
many helpful discussions, and to APC-Paris VII and Scuola Normale
Superiore for the kind hospitality extended to one or more of us.
The present research was supported in part by Scuola Normale
Superiore, by INFN and by the MIUR-PRIN contract 2009-KHZKRX. }

\appendix

\section{Linearized AdS Gravity}
\label{sec:lin GR}

We linearize the action for the AdS Gravity.
We first redefine the metric tensor as
\be
    G_{\sst MN}(x,z)=\frac{\eta_{\sst MN}+
    \kappa_{\sst MN}(x,z)}{z^{2}}\,,
\ee
and then expand the Hilbert-Einstein (HE) action $\mathcal{I}_{\sst HE}$
up to quadratic order in $\kappa_{\sst MN}$\,.
The latter is related to the notation used in this paper as
\be
    \varphi(x,z;U)=\tfrac12\,U^{\sst M}\,U^{\sst N}\,\kappa_{\sst MN}(x,z)\,.
\ee
We perform integrations by part to express $\mathcal{I}_{\sst HE}$ as
\be
    \mathcal{S}_{\sst HE}[G] = \mathcal{I}_{\sst \mathcal M}[\kappa]
    +\mathcal{I}^{\sst HE}_{\sst \partial \mathcal M}[\kappa] +\mathcal{O}(\kappa^{3})\,.
    \label{lin HE}
\ee
where $\mathcal{I}_{\sst \mathcal M}$ is the bulk action whose Lagrangian is
proportional to the EOM and $\mathcal{I}^{\sst HE}_{\sst \partial \mathcal M}$
the resulting boundary term. Explicitly the bulk action is given by
\be
    \mathcal{I}_{\sst \mathcal M}[\kappa]
    =-\frac14\int^{\infty}_{z_{B}} \frac{dz}{z^{d-1}}\int d^{d}x
    \left(\kappa^{\sst MN}\,F_{\sst MN}
    -\tfrac12\,\kappa^{\sst M}_{\sst M}\,F^{\sst N}_{\sst N}\right),
    \label{quad bulk}
\ee
where the indices ${\st M}, {\st N}$ are raised by the flat metric $\eta^{\sst MN}$
and the tensor $F_{\sst MN}$ is the linearization of
$-2\,(R_{\sst MN}+d\,G_{\sst MN})$\,.
The remaining part of  \eqref{lin HE}
is the boundary term $\mathcal{I}^{\sst HE}_{\sst \partial \mathcal M}$\,,
and it is given  by
\ba
    \mathcal{I}^{\sst HE}_{\sst \partial \mathcal M}[\kappa]=\int \frac{d^{d}x}{z_{B}^{d}}
    \!&\Big[\!&
    -z\,\partial_{z}\left(\kappa^{\mu}_{\mu}-\tfrac38\,\kappa_{\mu\nu}^{2}
    +\tfrac18\,(\kappa^{\mu}_{\mu})^{2}\right)+
    2\left(1+\tfrac12\,\kappa^{\mu}_{\mu}-\tfrac14\,\kappa_{\mu\nu}^{2}
    +\tfrac18\,(\kappa^{\mu}_{\mu})^{2}\right)\nn
    &&+\,\tfrac12\,\kappa_{dd}\,z\,\partial_{z}\,\kappa^{\mu}_{\mu}
    +\tfrac12\,z\left(\kappa^{\mu\nu}\,\partial_{\mu}\,\kappa_{\nu d}
    +2\,\kappa^{\mu d}\,\partial_{\mu}\,\kappa_{dd}+
    \kappa^{\mu d}\,\partial_{\mu}\,\kappa^{\nu}_{\nu}\right)\nn
    &&-\,d\left(\kappa_{dd}-\kappa_{\mu d}^{2}-\tfrac34\,\kappa_{dd}^{2}
    \right)
    -\tfrac14\,(d+1)\,\kappa_{dd}\,\kappa^{\nu}_{\nu}\,\Big]_{z=z_{B}}\,.
\ea
The HE action should be complemented by the York-Gibbons-Hawking (GH) term.
We linearize the latter as
$\mathcal{S}_{\sst GH}[G]= \mathcal{I}^{\sst GH}_{\sst\partial \mathcal M}[\kappa]
+\mathcal{O}(\kappa^{3})$ with
\ba
    \mathcal{I}^{\sst GH}_{\sst \partial \mathcal M}[\kappa]=
    \int \frac{d^{d}x}{z_{B}^{d}}\!&\Big[\!&
    z\,\partial_{z}\left(\kappa^{\mu}_{\mu}-\tfrac12\,\kappa_{\mu\nu}^{2}
    +\tfrac14\,(\kappa^{\mu}_{\mu})^{2}\right)
    -2d\left(1+\tfrac12\,\kappa^{\mu}_{\mu}-\tfrac14\,\kappa_{\mu\nu}^{2}
    +\tfrac18\,(\kappa^{\mu}_{\mu})^{2}\right) \nn
    &&-\,\tfrac12\,\kappa_{dd}\,z\,\partial_{z} \kappa^{\mu}_{\mu}
    +\kappa_{dd}\,z\,\partial^{\mu}\kappa_{\mu d}
    +d\left(\kappa_{dd}-\kappa_{\mu d}^{2}-\tfrac34\,\kappa_{dd}^{2}
    +\tfrac12\,\kappa_{dd}\,\kappa^{\mu}_{\mu}\right)
    \Big]_{z=z_{B}}.\hspace{30pt}
    \label{GH q}
\ea
Besides the GH term, we may include  an additional boundary term
proportional to the boundary volume
$V_{\sst \partial \mathcal M}[\gamma]=
\mathcal{I}_{\sst \partial \mathcal M}^{V}[\kappa]+
\mathcal{O}(\kappa^{3})$\,:
\be
    \mathcal{I}_{\sst \partial \mathcal M}^{V}[\kappa]
    =\int \frac{d^{d}x}{z_{B}^{d}}\,
    \Big[\,1+\tfrac12\,\kappa^{\mu}_{\mu}-\tfrac14\,\kappa_{\mu\nu}^{2}
    +\tfrac18\,(\kappa^{\mu}_{\mu})^{2}\,\Big]_{z=z_{B}}\,.
\ee
Finally we consider the following action for AdS Gravity\,:
\be
    \mathcal{S}_{\sst HE}[G]+\mathcal{S}_{\sst GH}[G]
    +c\,V_{\sst \partial \mathcal M}[\gamma]=
    \mathcal{I}_{\sst \mathcal M}[\kappa]
    +\mathcal{I}_{\sst \partial \mathcal M}[\kappa]+\mathcal{O}(\kappa^{3})\,,
    \label{gravity action}
\ee
where $\mathcal{I}_{\partial \mathcal M}$ is the
sum of at most quadratic part of different boundary terms:
\ba
    \label{bd term}
    && \mathcal{I}_{\sst \partial \mathcal M}[\kappa]=
    \mathcal{I}^{\sst HE}_{\sst \partial \mathcal M}[\kappa]
    +\mathcal{I}^{\sst GH}_{\sst \partial \mathcal M}[\kappa]
    +c\,\mathcal{I}_{\sst \partial \mathcal M}^{V}[\kappa] \nn
     &&=\, \int \frac{d^{d}x}{z_{B}^{d}}\,\Big[-\tfrac18\,z\,\partial_{z}\left(\kappa_{\mu\nu}^{2}
    -(\kappa^{\mu}_{\mu})^{2}\right)+
    [c-2(d-1)]\left(1+\tfrac12\,\kappa^{\mu}_{\mu}-\tfrac14\,\kappa_{\mu\nu}^{2}
    +\tfrac18\,(\kappa^{\mu}_{\mu})^{2}\right) \qquad\\
    &&\hspace{55pt}
    +\,\tfrac12\,\kappa^{\mu\nu}\,z\,\partial_{\mu}\kappa_{\nu d}
    -\tfrac 12\,\kappa^{\mu}_{\mu}\,z\,\partial^{\nu}\kappa_{\nu d}
    +\tfrac14\,(d-1)\,\kappa^{\mu}_{\mu}\,\kappa_{dd}\
    \Big]_{z=z_{B}}\,.\nonumber
    \label{bd spin2}
\ea
The choice $c=2\,(d-1)$ gives us a particularly simple result
where the constant term as well as the linear term in $\kappa$ disappear.

\section{Radial decomposition of the Fronsdal operator}
\label{sec:Fronsdal}

For an explicit analysis of the Fronsdal theory, it is useful to
decompose the Fronsdal operator as follows: \be
    \mathcal{F}=\sum_{n=0}^{2}\,\sum_{m=0}^{4}\,v^{n}\,
    \mathcal{F}^{n}_{m}\,\partial_{v}^{\, m}\,,
\ee where $\mathcal{F}^{n}_{m}$ are operators without $v$ and
$\partial_{v}$
 given by
\ba
    \mathcal{F}^{0}_{0}\e
     (z\,\partial_{z})^{2}-d\,z\,\partial_{z}-m_{s}^{2}-s
     +z^{2}\,\mathcal{F}_{M^{d}}-
    \tfrac12\,u^{2}\,(z\,\partial_{z}+u\cdot\partial_{u})\,\partial_{u}^{2}\,, \nn
    \mathcal{F}^{0}_{1}\e
    z\,u^{2}\,(\partial \cdot \partial_{u})
    +z\,(u\cdot \partial )\,(d-2-z\,\partial_{z}+
    u\cdot\partial_{u}-u^{2}\,\partial_{u}^{2})\,, \nn
    \mathcal{F}^{0}_{2}\e
    \tfrac12 \left[
    z^{2}\,(u\cdot \partial )^{2}+u^{2}\,
    (z\,\partial_{z}-2d-3\,u\cdot\partial_{u}+u^{2}\,\partial_{u}^{2})\right], \nn
    \mathcal{F}^{0}_{3}\e
    -z\,u^{2}\,(u\cdot\partial )\,,\qquad
    \mathcal{F}^{0}_{4}=\tfrac12\,(u^{2})^{2}\,,\nn
    \mathcal{F}^{1}_{0}\e
    z\,(u\cdot\partial \,\partial_{u}^{2}-\partial \cdot\partial_{u})
    (z\,\partial_{z}+u\cdot\partial_{u}-2)\, \nn
    \mathcal{F}^{1}_{1}\e
    -d-2\,s+3+(z\,\partial_{z}+u\cdot\partial_{u})\,(d-z\,\partial_{z}+u\cdot\partial_{u})
    -u^{2}\,(z\,\partial_{z}+u\cdot\partial_{u})\,\partial_{u}^{2}\,,\nn
    \mathcal{F}^{1}_{2}\e
    z\,(u\cdot\partial )\,(z\,\partial_{z}+u\cdot\partial_{u}+1)\,,\nn
    \mathcal{F}^{1}_{3}\e
    -u^{2}\,(z\,\partial_{z}+u\cdot\partial_{u}+1)\,,\qquad
    \mathcal{F}^{1}_{4}=0\,,\nn
    \mathcal{F}^{2}_{0}\e
    \tfrac12\,(z\,\partial_{z}+u\cdot\partial_{u})\,
    (z\,\partial_{z}+u\cdot\partial_{u}-2)\,\partial_{u}^{2}\,,\nn
    \mathcal{F}^{2}_{1}\e 0\,,
    \qquad \mathcal{F}^{2}_{2}=1+\tfrac12\,(z\,\partial_{z}+u\cdot\partial_{u})^{2}\,,
    \qquad \mathcal{F}^{2}_{3}=\mathcal{F}^{2}_{4}=0\,.
    \label{Fronsdal expl.}
\ea In \eqref{bd variat.}, the gauge variation of the bulk
Fronsdal action is given by two quantities:
$\partial_{v}\,\mathcal{F}\,\varphi^{\sst (s)}|_{v=0}$ and
$(\partial_{u}^{2}-\partial_{v}^{2})\,\mathcal{F}\,\varphi^{\sst
(s)}|_{v=0}$\,. Here, we compute them explicitly. First we rewrite
them in terms of the decomposed operators $\mathcal{F}^{n}_{m}$
and then use the results of \eqref{Fronsdal expl.}. Finally the
explicit expressions for these two quantities are \ba
    && \left(\partial_{v}\,\mathcal{F}\,\varphi^{\sst (s)}\right)\!(x,z;u,0) = \nn
    &&=\left[ \mathcal{F}^{1}_{0}-\mathcal{F}^{0}_{3}\,(\partial_{u}^{2})^{2}\right]
    \phi^{\sst (s)}(x,z;u)
    +\left[ \mathcal{F}^{0}_{0}+ \mathcal{F}^{1}_{1}-
    \mathcal{F}^{0}_{4}\,(\partial_{u}^{2})^{2}\right] \phi^{\sst (s-1)}(x,z;u) \nn
    && \quad +\,\left[ \mathcal{F}^{0}_{1}+ \mathcal{F}^{1}_{2}-
    2\,\mathcal{F}^{0}_{3}\,\partial_{u}^{2}\right]\phi^{\sst (s-2)}(x,z;u)
    +\left[ \mathcal{F}^{0}_{2}+ \mathcal{F}^{1}_{3}-
    2\,\mathcal{F}^{0}_{4}\,\partial_{u}^{2}\right]\phi^{\sst (s-3)}(x,z;u)
    \nn
    &&=
    \left[ z\, ( u\cdot \partial \,\partial_{u}^{2}-\partial \cdot\partial_{u})
    (z\,\partial_{z}+s-2)
    +z\,u\cdot\partial \,u^{2}\,(\partial_{u}^{2})^{2}\right] \phi^{\sst (s)}(x,z;u) \nn
    &&\quad +\,\left[ z^{2}\,\mathcal{F}_{M^{d}}-\tfrac32\,u^{2}\, (z\,\partial_{z}+s-3)\,
    \partial_{u}^{2}-\tfrac12\,(u^{2})^{2}\,(\partial_{u}^{2})^{2}\right]\phi^{\sst (s-1)}(x,z;u) \nn
    &&\quad +\,\left[ (d+2s-5)\,z\,u\cdot\partial + z\,u^{2}
    \left(\partial \cdot\partial_{u}+u\cdot\partial \,\partial_{u}^{2}\right)
     \right] \phi^{\sst (s-2)}(x,z;u)\nn
    &&\quad +\,\left[\tfrac12\,z^{2}\,(u\cdot\partial )^{2}-
    \tfrac12\,u^{2} \left(z\,\partial_{z}+2\,d+5\,s-13+u^{2}\,\partial_{u}^{2}\right)\right]
    \phi^{\sst (s-3)}(x,z;u)\,,
    \label{F s-1}
\ea and \ba
    && \left( (\partial_{u}^{2}-\partial_{v}^{2})\,
    \mathcal{F}\,\varphi^{\sst (s)}\right)\!(x,z;u,0)= \nn
    && =\left[
    \partial_{u}^{2} \left(
    \mathcal{F}^{0}_{0}-\mathcal{F}^{0}_{4}\,(\partial_{u}^{2})^{2}\right)
    -2\,\mathcal{F}^{2}_{0}+
    \left(\mathcal{F}^{0}_{2}+2\,\mathcal{F}^{1}_{3}-
    2\,\mathcal{F}^{0}_{4}\,\partial^{2}_{u}\right)(\partial_{u}^{2})^{2}
    \right] \phi^{\sst (s)}(x,z;u) \nn
    && \quad +\,
    \left[  \partial_{u}^{2}
    \mathcal{F}^{0}_{1}-2\,\mathcal{F}^{1}_{0}+
    \mathcal{F}^{0}_{3}\,(\partial_{u}^{2})^{2}
    \right] \phi^{\sst (s-1)}(x,z;u) \nn
    && \quad +\,
    \left[ \partial_{u}^{2} \left(
    \mathcal{F}^{0}_{2}-2\,\mathcal{F}^{0}_{4}\,\partial_{u}^{2}\right)
    -\mathcal{F}^{0}_{0}-2\,\mathcal{F}^{1}_{1}-2\,\mathcal{F}^{2}_{2}+
    \left(2\,\mathcal{F}^{0}_{2}+4\,\mathcal{F}^{1}_{3}
    -3\,\mathcal{F}^{0}_{4}\,\partial_{u}^{2}\right) \partial_{u}^{2}
    \right] \phi^{\sst (s-2)}(x,z;u) \nn
    && \quad +\,
    \left[ \partial_{u}^{2} \mathcal{F}^{0}_{3}
    -\mathcal{F}^{0}_{1}    -2\,\mathcal{F}^{1}_{2}+
    2\,\mathcal{F}^{0}_{3}\,\partial^{2}_{u}\right] \phi^{\sst (s-3)}(x,z;u) \nn
    && =
    \Big[  -
    2\,(d+2\,s-5)\,(z\,\partial_{z}+s-2)\,\partial_{u}^{2}
     - u^{2}\left( 2\,z\,\partial_{z}+3d+8s-25+u^{2}\partial_{u}^{2}\right)
    (\partial_{u}^{2})^{2} \nn
    && \qquad\qquad +\,
    z^{2} \left[ 2\left(\Box\,\partial_{u}^{2}-
    (\partial_{u}\cdot\partial )^{2}\right)
    +u\cdot\partial \left(\partial_{u}\cdot \partial +
    u\cdot\partial \,\partial_{u}^{2}\right) \partial_{u}^{2}\right]
    \Big]\, \phi^{\sst (s)}(x,z;u) \nn
    &&\quad +\,
    z\,\Big[ 4\,(d+2\,s-5)\,\partial_{u}\cdot\partial -
    (3\,z\,\partial_{z}+d+5\,s-11)\,u\cdot\partial \,\partial_{u}^{2} \nn
    && \qquad\qquad
    -\,u^{2}\,\partial_{u}\cdot\partial \,\partial_{u}^{2}
    -2\,u^{2}\,u\cdot\partial \,(\partial_{u}^{2})^{2}\Big]
    \phi^{\sst (s-1)}(x,z;u) \nn
    &&\quad +\,
    \Big[ z^{2}\,u\cdot\partial  \left( 3\,\partial_{u}\cdot\partial
    +u\cdot\partial \,\partial_{u}^{2}\right)
    -(u^{2})^{2}(\partial_{u}^{2})^{2}\nn
    &&\qquad\qquad-\,
    (5\,d+10\,s-29)\,u^{2}\,\partial_{u}^{2}
    -2\,(d+2\,s-4)\,(d+2\,s-5)\Big]\,\phi^{\sst (s-2)}(x,z;u) \nn
    &&\quad-z\,
    \Big[ u\cdot\partial \,\left(
    z\,\partial_{z}+3\,d+7\,s-17+2\,u^{2}\,\partial_{u}^{2}\right)
    +3\,u^{2}\,\partial_{u}\cdot\partial \Big]\,\phi^{\sst (s-3)}(x,z;u)\,.
    \label{F s-2}
\ea

\section{Useful identities}
\label{sec: IDs}

In this section, we summarize the mathematical identities that has been used in the paper.
First, the $\mathsf P$-operators
introduced for the construction of boundary actions satisfy
\ba
\label{P id1}
    && \mathsf{P}_{\rm e}-\mathsf{P}_{\rm o}=u^{2}\,\mathsf{P}'_{\rm o}\,\partial_{u}^{2}\,,\qquad
    \mathsf{P}_{\rm e}+\mathsf{P}'_{\rm e}=u^{2}\,\mathsf{P}_{\rm o}\,\partial_{u}^{2}\,,\nn
    &&\left[3\left(d+2\,u\cdot\partial_{u}+1\right)+u^{2}\,\partial_{u}^{2}\right] \mathsf{P}'_{\rm o}
    =\left(d+2\,u\cdot\partial_{u}+1\right) \mathsf{P}_{\rm o}\,,
\ea
and
\ba
\label{P id2}
    &\partial_{u}\cdot\partial \,\mathsf{P}_{\rm e}'
    =\mathsf{P}_{\rm e}\left(u\cdot\partial \,\partial_{u}^{2}-\partial_{u}\cdot\partial \right),
    \qquad
    &\partial_{u}^{2}\,\mathsf{P}_{\rm e}'=(d+2\,u\cdot\partial_{u}-1)\,\mathsf{P}_{\rm e}\,\partial_{u}^{2}\,,\nn
    &\partial_{u}\cdot\partial \,\mathsf{P}_{\rm e}
    =\mathsf{P}_{\rm o}\left(u\cdot\partial \,\partial_{u}^{2}+\partial_{u}\cdot\partial \right),
    \qquad
    &\partial_{u}^{2}\,\mathsf{P}_{\rm e}=(d+2\,u\cdot\partial_{u}+1)\,\mathsf{P}_{\rm o}\,\partial_{u}^{2}\,,\nn
    &\ \ \partial_{u}\cdot\partial \,\mathsf{P}_{\rm o}
    =\mathsf{P}_{\rm o}'\left(u\cdot\partial \,\partial_{u}^{2}+3\,\partial_{u}\cdot\partial \right),
    \qquad
    &\partial_{u}^{2}\,\mathsf{P}_{\rm o}=(d+2\,u\cdot\partial_{u}+3)\,\mathsf{P}'_{\rm o}\,\partial_{u}^{2}\,,
\ea

The Bessel-like function $U_{\nu}(z)$  appears in the solutions of
the Fronsdal equation. It is a
 related to the
modified Bessel function of second kind: \be
    U_{\nu}(z):=\frac{2}{\Gamma(\nu)}
    \left(\frac z2\right)^{\nu}\,K_{\nu}(z)=1+\mathcal O(z^{2})\,,
    \label{U def}
\ee and it satisfies the following Bessel-like differential
equation: \be
    \left[ (z\,\partial_{z}-\nu)^{2}-a\,z^{2}-\nu^{2}
    \right] U_{\nu}(a\,z)=0\,.
\ee In order to study the finite part or the logarithmically
divergent part of the on-shell action, we need the series
expansion of this function: \be
    U_{\nu}(z)
    = \sum_{k=0}^{n_{\nu}} \frac{\left(\tfrac z2\right)^{2k}}{(1-\nu)_{k}\,k!}
     + \frac{\left(\tfrac z2\right)^{2\nu}}
    {\Gamma(\nu)\,\Gamma(\nu+1)}
     \sum_{k=0}^{\infty}\, u_{\nu,k}(z)\,\frac{\left(\tfrac z2\right)^{2k}}{(1+\nu)_{k}\,k!}\,,
     \label{U exp}
\ee where $n_{\nu}$ is given by \be
    n_{\nu}=\left\{
    \begin{array}{cc}
    \infty \qquad & [\nu-\tfrac12 \in \mathbb N] \\
    \nu-1 \qquad & [\nu \in \mathbb N]
    \end{array}
    \right.,
\ee
and $u_{\nu,k}(z)$ by \be
    u_{\nu,k}(z)=
    \left\{
    \begin{array}{cc}
     \pi\,(-1)^{\nu+\frac12} \qquad
    & [\nu-\tfrac12 \in \mathbb N] \\
    (-1)^{\nu}\left[\psi(k+1)+\psi(k+\nu+1)-
     2\,\ln\!\left(\frac z2\right)\right] \qquad
     & [\nu \in \mathbb N]
     \end{array}
     \right..
\ee The function $U_{\nu}(z)$ enjoys several identities that can
be proven from the identities of the modified Bessel function
$K_{\nu}$\,. For example, it satisfies the following recurrence
identity: \be
    U_{\nu}(z)-U_{\nu-1}(z)=\tfrac{z^{2}}{4(\nu-1)(\nu-2)}\,U_{\nu-2}(z)\,.
    \label{U id1}
\ee The derivatives of $U_{\nu}$ can be again expressed in terms
of $U_{\nu'}$ with a neighboring index:
\be
    z\,\partial_{z}\,U_{\nu}(z)
    =-\tfrac{z^{2}}{2(\nu-1)}\,U_{\nu-1}(z)\,.
        \label{U id}
\ee

\section{Gauge variation of the bulk action and the boundary action}
\label{sec: gauge var}

The variation of the bulk action under $\epsilon^{\sst(s-2)}$ gives
\be
    \delta_{\epsilon^{(s-2)}}\,\mathcal I_{\sst \mathcal M}=
    \tfrac{z_{B}^{-d}}{2}\,
    \ppd{\epsilon^{\sst (s-2)}}{\mathsf P_{\rm o}\,\tfrac12\,(\partial_{v}^{2}-\partial_{u}^{2})\,\mathcal F\,\varphi}_{z_{B}}
    =\tfrac{z_{B}^{-d}}2\,\ppd{\epsilon^{\sst (s-2)}}{J_{\sst\mathcal M}}_{z_{B}}\,,
\ee
where $J_{\sst \mathcal M}$ is given by
\ba
    J_{\sst \mathcal M} &=&\mathsf P_{\rm o}\,\Big[
    (d+2s-5+u^{2}\,\partial_{u}^{2})\,\partial_{u}^{2}\,\chi^{\sst (s)}+z\,u\cdot\partial\,\chi^{\sst (s-3)}
    +(d+2s-4+u^{2}\,\partial_{u}^{2})\,\zeta^{\sst (s-2)} \nn
    && \qquad -\, \tfrac12\,z^{2}\,(\partial_{u}^{2}\,B^{\sst (s)}-B^{\sst (s-2)})
    -\tfrac12\left(3d+6s-11+3\,u^{2}\,\partial_{u}^{2}\right)u^{2}\,\partial_{u}^{2}\,\phi^{\sst (s)}\nn
    &&\qquad +\, \tfrac12\,(d+2s-1+u^{2}\,\partial_{u}^{2})\,u^{2}\,\partial_{u}^{2}\,\phi^{\sst (s-2)}
    +z\,u^{2}\,\partial_{u}\cdot\partial\,\psi^{\sst (s-3)}\,\Big]\,,\ea
where we used the gauge invariants:
 \ba
    && {B}^{\sst (s)}:={\cal F}_{M^{d}}\,\phi^{\sst (s)}+\tfrac12\,(u\cdot\partial)^2\,\phi^{\sst (s-2)}\,, \nn
   && {B}^{\sst (s-2)}:= {\cal F}_{M^{d}}\,\phi^{\sst (s-2)}-\tfrac12\,
    (u\cdot\partial)^2\left[
    (\partial_u^2)^2\,\phi^{\sst (s)}+2\,\partial_u^2\,\phi^{\sst (s-2)}\right].
\ea
The variation of the boundary terms \eqref{boundary}
under $\epsilon^{\sst (s-2)}$ gives
 \be
    \delta_{\epsilon^{\sst (s-2)}}\,\mathcal I_{\sst \partial\mathcal M}=
    \tfrac{z_{B}^{-d}}2\,\ppd{\epsilon^{\sst (s-2)}}{J_{\sst \partial\mathcal M }}_{z_{B}}\,,
\ee
where $J_{\sst \partial\mathcal M}$ is given by
\ba
    J_{\sst \partial\mathcal M} &=& \mathsf P_{\rm o}\,\Big[
    -(d+2s-5+u^{2}\,\partial_{u}^{2})\,\partial_{u}^{2}\,\chi^{\sst (s)}-z\,u\cdot\partial\,\chi^{\sst (s-3)}
    +(d+s-3)\,z\,u\cdot\partial\,\psi^{\sst (s-3)}\nn
    &&\qquad +\, z^{2}\,u\cdot\partial\,(u\cdot\partial\,\partial_{u}^{2}+3\,\partial_{u}\cdot\partial)(\xi^{\sst (s-2)}+\mathsf T\,\zeta^{\sst (s-2)}) \nn
    && \qquad -\, z^{2} \left[ \Box\,\partial_{u}^{2}-(\partial_{u}\cdot\partial)^{2}+2\,u\cdot\partial\,\partial_{u}\cdot\partial\,\partial_{u}^{2}
    +(u\cdot\partial)^{2}\,(\partial_{u}^{2})^{2} \right] \phi^{\sst (s)} \nn
    && \qquad -\, (s-2+\tfrac12\,\partial_{u}^{2}\,u^{2}+\tfrac12\,u^{2}\,\partial_{u}^{2})\,\zeta^{\sst (s-2)} \nn
    && \qquad +\left[
    (2(s-2)+u^{2}\,\partial_{u}^{2})(3(d+2s-5)+u^{2}\,\partial_{u}^{2})-2\,\partial_{u}^{2}\,(u^{2})^{2}\,\partial_{u}^{2}\right]
    (\xi^{\sst (s-2)}+\mathsf T\,\zeta^{\sst (s-2)}) \Big]\nn
    &&-\,(d+2s-5)\left[
    (3(s-2)+u^{2}\,\partial_{u}^{2})\,\mathsf P_{\rm e}\,\partial_{u}^{2}\,\phi^{\sst (s)}+
    z\,u\cdot\partial\, \mathsf P_{\rm o}\,\psi^{\sst (s-3)}\right].
\ea

\section{Counterterms}
\label{sec: counter}

The $\epsilon^{\sst (s-2)}$-variations of the counterterm
$\mathcal I^{\sst\,2\,{\rm(c.t.)}}_{\sst\partial\mathcal M}$ \eqref{c.t. 2}
can be obtained by noticing first that it
depends on $\beta^{\sst (s-3)}$ through $\gamma^{\sst (s-2)}$
\be \gamma^{\sst (s-2)}=z\,u\cdot\partial\,\beta^{\sst (s-3)}\,,\ee so that
\ba
   \tfrac{z_{B}^{d}}{c}\,\mathcal I^{\sst\,2\,{\rm(c.t.)}}_{\sst\partial\mathcal M}\e
  \ppd{\phi^{\sst (s)}}{G^{\sst (s)}}_{z_{B}}-\ppd{\gamma^{\sst (s-2)}}{{A}^{\sst (s-2)}}_{z_{B}}\nn
    \e \ppd{\phi^{\sst(s)}}{\mathcal G_{\sst \phi\phi}\,\phi^{\sst (s)}}_{z_{B}}
    +\ppd{\phi^{\sst (s)}}{\mathcal G_{\sst \phi\gamma}\,\gamma^{\sst (s-2)}}_{z_{B}} \nn
    &&+\,\ppd{\gamma^{\sst (s-2)}}{\mathcal G_{\sst \gamma\phi}\,\phi^{\sst (s)}}_{z_{B}}
    +\ppd{\gamma^{\sst (s-2)}}{\mathcal G_{\sst \gamma\gamma}\,\gamma^{\sst (s-2)}}_{z_{B}} \,,
\ea
with
\ba \mathcal G_{\sst \phi\phi} \e
z^{2}\left[\Box-u\cdot\partial\,\partial_u\cdot\partial
+\tfrac{1}{2}\,(u\cdot\partial)^2\,\partial_u^2 -\tfrac{1}{2}\,u^2
\left(\Box\,\partial_u^2-(\partial_u\cdot\partial)^2-\tfrac{1}{2}
u\cdot\partial\,\partial_u\cdot\partial\,\partial_u^2\right)\right],\nn
\mathcal G_{\sst \phi\gamma}\e \tfrac{z^{2}}{2}\left[u^2
\left(\Box+\tfrac{1}{2}\,u\cdot\partial\,\partial_u\cdot\partial\right)
-(u\cdot\partial)^2\right],\nn
\mathcal G_{\sst \gamma\phi} \e \mathcal G_{\sst \phi\gamma}^{\dagger}\,,
\qquad \mathcal G_{\sst \gamma\gamma}=
z^{2}\left(\Box+\tfrac{1}{2}\,u\cdot\partial\,\partial_u\cdot\partial\right).
\ea
Notice the selfadjoint-ness of the above operators:
\be
\mathcal G_{ij}^{\dagger}=\mathcal G_{ji}\,,\ee so that under an arbitrary variation,
we have
\be \delta \, \mathcal I^{\sst\,2\,{\rm(c.t.)}}_{\sst\partial\mathcal M}
=2\,c\, z_{B}^{-d}\,
\Big[\,\ppd{\delta\,\phi^{\sst (s)}}{G^{\sst (s-3)}}_{z_{B}}\!+\,
\ppd{\delta\,\beta^{\sst (s-3)}}{K^{\sst (s-3)}}_{z_{B}}\,\Big]\,.
\label{var c.t. 2}
\ee
Using the $\epsilon^{\sst (s-2)}$-variations:
\be \delta_{\epsilon^{\sst (s-2)}}\,\phi^{\sst (s)}=
-u^2\,\epsilon^{\sst (s-2)}\,, \qquad \delta_{\epsilon^{\sst (s-2)}}\,\beta^{\sst (s-3)}=
z\,\mathsf T\,(u\cdot\partial\,\partial_u^2+3\,\partial_u\cdot\partial)\,\epsilon^{\sst (s-2)}\,,
\ee
we obtain \eqref{var c.t. 2}.

\section{Traceless and transverse projection }
\label{sec: TT app}

The traceless and transverse part of field $h^{\sst (s)}$
can be uniquely determined
from the gauge transformation:
\be
    \delta\,h^{\sst (s)}=u\cdot\partial\,\lambda^{\sst (s-1)}-u^{2}\,\sigma^{\sst (s-2)}\,.
    \label{free gauge}
\ee
Without loss of generality, we assume that the gauge
parameter $\lambda^{\sst (s-1)}$ is traceless: $\lambda^{\sst (s-1)}=
\bar\lambda^{\sst (s-1)}$\,,  since its trace part
overlaps the divergent part of $\lambda^{\sst (s-2)}$\,. Then, the trace
of $\sigma^{\sst (s-2)}$ is uniquely fixed by the double trace part of
$h^{\sst (s)}$ via \be
    (\partial_{u}^{2})^{2}\,h^{\sst (s)}
    +\left[4(d+2s-6)+u^{2}\,\partial_{u}^{2}\right] \partial_{u}^{2}\,\sigma^{\sst (s-2)}=0\,.
\ee Then, we have a doubly traceless field $\bar{\bar h}^{\sst
(s)}$ with two traceless parameters $\bar \lambda^{\sst (s-1)}$ and
$\bar\sigma^{\sst (s-2)}$\,. Now, we use the equations of the
traceless and transverse gauge conditions. First, the
traceless gauge condition determines $\bar \sigma^{\sst (s-2)}$ in
terms of $\bar{\bar h}^{\sst (s)}$ and $\bar \lambda^{\sst (s-1)}$ as
\be
    \bar \sigma^{\sst (s-2)}=-\tfrac1{2(d+2s-4)}\,
    \left[2\,\partial_{u}\cdot\partial \,\bar \lambda^{\sst (s-1)}+
    \partial_{u}^{2}\,\bar{\bar h}^{\sst (s)}
    \right].
\ee By plugging in this solution into the transverse
condition, we get the following equation for $\bar \lambda^{\sst
(s-1)}$\,: \ba
    &&\left[\Box+\tfrac{d+2s-6}{d+2s-4}\,u\cdot\partial \,
    \partial_{u}\cdot\partial -\tfrac1{d+2s-4}\,u^{2}\,(\partial_{u}\cdot\partial )^{2}
    \right] \bar \lambda^{\sst (s-1)} \nn
    &&\quad +\left[\left(1-\tfrac{u^{2}\,\partial_{u}^{2}}{2(d+2s-4)}\right)
    \partial_{u}\cdot\partial -\tfrac{u\cdot\partial \,\partial_{u}^{2}}{d+2s-4}
    \right] \bar{\bar h}^{\sst (s)} =0\,.
\ea By acting a $r$-ple divergence of the above equation, we get
$(\partial_{u}\cdot\partial )^{r}\, \bar \lambda^{\sst (s-1)}$ in terms
of $(\partial_{u}\cdot\partial )^{r+1}\, \bar \lambda^{\sst (s-1)}$\,,
$(\partial_{u}\cdot\partial )^{r+2}\, \bar \lambda^{\sst (s-1)}$ and
$\bar{\bar h}^{\sst (s)}$\,:
\ba
    (\partial_{u}\cdot\partial )^{r} \bar \lambda^{\sst (s-1)}\e
    \tfrac1{(r+1)(d+2s-r-4)}\,
    \Big[ (d+2(s-r)-6)\,\tfrac{u\cdot\partial }{\Box}\,
    (\partial_{u}\cdot\partial )^{r+1} \bar \lambda^{\sst (s-1)}
    -\tfrac{u^{2}}{\Box}\,(\partial_{u}\cdot\partial )^{r+2}\,
    \bar \lambda^{\sst (s-1)}\Big] \nn
    &&
    +\,\tfrac{d+2s-4}{(r+1)(d+2s-r-4)}\,
    \tfrac{(\partial_{u}\cdot\partial )^{r}}{\Box}
    \left[\left(1-\tfrac{u^{2}\,\partial_{u}^{2}}{2(d+2s-4)}\right)
    \partial_{u}\cdot\partial -\tfrac{u\cdot\partial \,\partial_{u}^{2}}{d+2s-4}
    \right] \bar{\bar h}^{\sst (s)}\,.
\ea Thus, one can solve iteratively the multiple divergences of
$\bar\lambda^{\sst (s-1)}$\,, and finally $\bar\lambda^{\sst (s-1)}$ itself.
This procedure is unique, and finally the gauge parameters
$\lambda^{\sst (s-1)}$ and $\sigma^{\sst (s-2)}$ are uniquely determined by
the traceless and transverse gauge conditions on $h^{\sst
(s)}$\,.

Second method:
 In fact it is possible to find a closed expression
of $h^{\sst (s)}_{\sst\rm TT}$. First, we project $h^{\sst (s)}$
on its transverse part $h^{\sst (s)}_{\sst\rm T}$ as \be
h^{\sst(s)}_{\sst\rm T}(x,u)=\tfrac{1}{s!}\,\Big(u\cdot\partial_w-
\tfrac{u\cdot\partial\,\partial_w\cdot\partial}{\Box}\Big)^s\,h^{\sst
(s)}(x,w)= \left(\tfrac{\partial_{\tilde
u}\cdot\partial}\Box\right)^{s}\,\mathscr R^{\sst (s)}(u,\tilde
u)\,, \label{proj t} \ee where $\mathscr R^{\sst (s)}$ is the
deWit-Freedman curvature \cite{deWit:1979pe}:
\be
\mathscr R^{\sst (s)}(u,\tilde u)=\tfrac1{s!} \left( u\cdot\partial_{w}\,\tilde
u\cdot\partial-u\cdot\partial\,\tilde
u\cdot\partial_{w}\right)^{s}\, h^{\sst (s)}(x,w)\,.
\label{curv s}
 \ee
It is also possible to express $h^{\sst (s)}_{\sst\rm T}$ as
\be
    h^{\sst (s)}_{\sst\rm T}(x,u)=
    \sum_{n=0}^{s}\,\tfrac{(-1)^{n}}{n!}\,
    \tfrac{(u\cdot\partial)^{n}(\partial_{u}\cdot\partial)^{n}}{\Box^{n}}\,
    h^{\sst (s)}(x,u)=\
    :e^{-\frac{u\cdot\partial\,\partial_{u}\cdot\partial}\Box}:\,
    h^{\sst (s)}(x,u)\,,
\ee
where :\,: is the normal product.
The projection to $h^{\sst (s)}_{\sst\rm TT}$ can be obtained by using
the identity: \be
\left[\,u^2-\tfrac{(u\cdot\partial)^2}{\Box}\,,\,\partial_u\!\cdot\partial\,\right]=0\,,
\ee and by taking successive traces of $h^{\sst (s)}_{\sst\rm T}$
as \be h^{\sst (s)}_{\sst\rm TT}(x,u)=\sum_n a_n\,\Big(
u^2-\tfrac{(u\cdot\partial)^2}{\Box}\Big)^n\,(\partial_u^2)^n\,
h^{\sst (s)}_{\sst\rm T}(x,u)\,. \ee Finally, by requiring the
traceless condition, the constants $a_n$ are determined as \be
a_n= \frac{1}{4^n\,n!\,(-\frac{d+2s-5}2)_n}\,, \ee and replacing
$h^{\sst (s)}_{\sst\rm T}$ by its expression and after some
algebra, we get \ba h^{\sst (s)}_{\sst\rm TT}(x,u) \e
\sum_{n=0}^{\infty}\,
\frac{1}{4^n\,n!\,(s-2n)!\,(\frac{1}{2}-\nu)_n}
\left(u^2-\tfrac{(u\cdot\partial)^2}{\Box}\right)^n \times \nn &&
\qquad \times
\left(\partial_w^2-\tfrac{(\partial_w\cdot\partial)^2}{\Box}\right)^n
\Big(u\cdot\partial_w-\tfrac{u\cdot\partial\partial_w\cdot\partial}{\Box}\Big)^{s-2n}\,h^{\sst
(s)}(x,w)\,. \ea The projector $P^{\sst (s)}_{\sst\rm TT}$ can be
deduced as
 \be
 \mathcal P^{\sst (s)}_{\sst\rm TT} =
\sum_{n=0}^{[s/2]}\,
\frac{1}{4^n\,n!\,(-\frac{d+2s-5}2)_n}
 \left(u^2-\tfrac{(u\cdot\partial)^2}{\Box}\right)^n
:e^{-\frac{u\cdot\partial\, \partial_u\cdot\partial}{\Box}}:
\left(\partial_u^2-\tfrac{(\partial_u\cdot\partial)^2}{\Box}\right)^n\,
\label{P tt}. \ee
 Notice that
$\Box^s\,\mathcal P^{\sst (s)}_{\sst\rm TT}$ is a local operator.

Finally, the uniqueness of the gauge-fixing procedure under
\eqref{free gauge} enables us to easily obtain the contraction of
the spin $s$ Weyl tensors $\mathscr C^{\sst (s)}$.\footnote{The
spin $s$ Weyl tensor is the mixed symmetry traceless projection of
the  curvature $\mathscr R^{\sst (s)}(u,\tilde u)$} Since the
latter is invariant under \eqref{free gauge}, we first gauge fix
$h^{\sst (s)}$ to $h^{\sst (s)}_{\sst\rm TT}$\,. Then, taking into
account that $\mathscr C^{\sst (s)}$ contains $s$ derivatives
acting on $h^{\sst (s)}_{\sst\rm TT}$, there exists only one
possible expression for $\ppd{\mathscr C^{\sst (s)}}{\mathscr
C^{\sst (s)}}$ expressed in terms of $h^{\sst (s)}_{\sst\rm TT}$
which is $\ppd{h^{\sst (s)}_{\sst\rm TT}}{(-\Box)^{s}\,h^{\sst
(s)}_{\sst\rm TT}}$  up to an overall constant.
%    In the text this constant
%    was absorbed in the definition of the scalar product acting on
%    mixed tensors or alternatively in the definition of $\mathscr C^{\sst
%    (s)}$.

\bibliographystyle{JHEP}
\bibliography{ref_duality}

\end{document}